\documentclass[a4paper,11pt,amsmath,amssymb]{article} 
\usepackage{jheppub}
\usepackage{times}
\usepackage{xcolor}
\usepackage{graphicx} 
\usepackage{microtype}
\usepackage{float}
\usepackage{subcaption}
\usepackage{ulem}
\usepackage{soul}

\def\beq{\begin{equation}}
\def\eeq{\end{equation}}
\def\bea{\begin{eqnarray}}
\def\eea{\end{eqnarray}}
\def\beqn{\begin{eqnarray}} 
\def\eeqn{\end{eqnarray}}
\def\nn{\nonumber}
\def\Eq#1{Eq.~(\ref{#1})}

\def\qon#1{q_{#1,0}^{(+)}}

\def\bra#1{\langle{#1}|}
\def\ket#1{|{#1}\rangle}
\def\id{\boldsymbol I}
\def\qb{\mathbf{q}}

\def\v{{\rm V}}

\def\ii{\imath 0}

\pdfoutput=1

\newcommand{\valencia}{Instituto de F\'{\i}sica Corpuscular, Universitat de Val\`{e}ncia -- Consejo Superior de Investigaciones Cient\'{\i}ficas, 
Parc Cient\'{\i}fic, E-46980 Paterna, Valencia, Spain.}
\newcommand{\culiacanA}{Facultad de Ciencias F\'{\i}sico-Matem\'aticas,
Universidad Aut\'onoma de Sinaloa, Ciudad Universitaria, CP 80000 Culiac\'an, Mexico.}
\newcommand{\culiacanV}{Facultad de Ciencias de la Tierra y el Espacio,
Universidad Aut\'onoma de Sinaloa, Ciudad Universitaria, CP 80000 Culiac\'an, Mexico.}
\newcommand{\berlin}{Deutsches Elektronen-Synchrotron DESY, Platanenallee 6, 15738 Zeuthen, Germany.}

\begin{document}
\normalem 

\title{Quantum algorithm for Feynman loop integrals}

\author[a,b,c]{Selomit Ram\'{\i}rez-Uribe,} 
\author[a]{Andr\'es E. Renter\'{\i}a-Olivo,}
\author[a]{Germ\'an Rodrigo,}
\author[d,a]{German F. R. Sborlini}
\author[a]{and Luiz Vale Silva} 

\affiliation[a]{\valencia}
\affiliation[b]{\culiacanA}
\affiliation[c]{\culiacanV}
\affiliation[d]{\berlin}

\emailAdd{norma.selomit.ramirez@ific.uv.es}
\emailAdd{andres.renteria@ific.uv.es}
\emailAdd{german.rodrigo@csic.es}
\emailAdd{german.sborlini@desy.de}
\emailAdd{luizva@ific.uv.es}

\preprint{IFIC/21-15, DESY 21-067}

\abstract{We present a novel benchmark application of a quantum algorithm to 
Feynman loop integrals. The two on-shell states of a Feynman propagator are 
identified with the two states of a qubit and a quantum algorithm is used to 
unfold the causal singular configurations of multiloop Feynman diagrams. To 
identify such configurations, we exploit Grover's algorithm for querying multiple 
solutions over unstructured datasets, which presents a quadratic speed-up over classical algorithms 
when the number of solutions is much smaller than the number of possible configurations. A 
suitable modification is introduced to deal with topologies in which the number of causal states to be identified is nearly half of the total number of states. The output of the quantum algorithm in \emph{IBM Quantum} and \emph{QUTE Testbed} simulators is used to bootstrap the
causal representation in the loop-tree duality of representative multiloop topologies. The 
algorithm may also find application and interest in graph theory to solve problems involving 
directed acyclic graphs.}

\setcounter{page}{1}
\maketitle

\section{Introduction and motivation}
\label{sec:Introduction}
Quantum algorithms~\cite{Feynman:1981tf} are a very promising avenue for solving specific problems that become too complex or even intractable 
for classical computers because they scale either exponentially or superpolynomially. They are particularly well suited to solve those problems for which the quantum principles of superposition and entanglement can be exploited to gain a speed-up advantage over the counterpart classical algorithms. These are, for example, the well-known cases of database querying~\cite{Grover:1997fa} and factoring integers into primes~\cite{Shor:1994jg}. Other recent applications are related to the enhanced capabilities of quantum systems for minimizing Hamiltonians \cite{APOLLONI1989233,PhysRevE.58.5355} which lead to a wide range of applications in optimization problems. For instance, this framework has been used in quantum
chemistry~\cite{Liu:2020eoa}, nuclear physics~\cite{Lynn:2019rdt,Holland:2019zju}, and also finance, such as portfolio optimization~\cite{Or_s_2019}.

The stringent demands that high-energy physics will meet in the coming Run 3 of the CERN's Large Hadron Collider (LHC)~\cite{Strategy:2019vxc}, the posterior high-luminosity phase~\cite{Gianotti:2002xx}, and the planned future colliders~\cite{Abada:2019lih,Djouadi:2007ik,Roloff:2018dqu,CEPCStudyGroup:2018ghi} motivate exploring new technologies. An interesting prospective avenue is quantum algorithms, which have recently started to come under the spotlight of the particle physics community. Recent applications include:
the speed up of jet clustering algorithms~\cite{Wei:2019rqy,Pires:2021fka,Pires:2020urc},
jet quenching~\cite{Barata:2021yri},
determination of parton densities~\cite{Perez-Salinas:2020nem}, 
simulation of parton showers~\cite{Bauer:2019qxa,Bauer:2021gup,deJong:2020tvx},
heavy-ion collisions~\cite{deJong:2020tvx},
quantum machine learning~\cite{Guan:2020bdl, Wu:2020cye,Trenti:2020ceh}
and lattice gauge theories~\cite{Jordan:2011ne,Banuls:2019bmf,Zohar:2015hwa,Byrnes:2005qx,Ferguson:2020qyf,Kan:2021nyu}.

One of the core bottlenecks in high-energy physics concerns the theoretical evaluation of quantum fluctuations at higher orders in the perturbative expansion by means of multiloop Feynman diagrams and the combination of all the ingredients contributing to a physical observable to provide accurate theoretical predictions beyond the second order or next-to-leading order (NLO).
Impressive advances have been achieved in recent years in this field.
For a very complete review of the current available frameworks, we refer the interested reader to Ref.~\cite{Heinrich:2020ybq}. They involve analytical, fully numerical and semi-analytical approaches for the evaluation of multiloop Feynman integrals, including sector decomposition~\cite{Binoth:2000ps,Smirnov:2008py,Carter:2010hi,Borowka:2017idc}, Mellin-Barnes transformation~\cite{Blumlein:2000hw,Anastasiou:2005cb,Bierenbaum:2006mq,Gluza:2007rt,Freitas:2010nx,Dubovyk:2016ocz}, algebraic reduction of integrands~\cite{Mastrolia:2011pr,Badger:2012dp,Zhang:2012ce,Mastrolia:2012an,Mastrolia:2012wf,Ita:2015tya,Mastrolia:2016dhn,Ossola:2006us}, integration-by-parts identities~\cite{Chetyrkin:1981qh,Laporta:2001dd}, semi-numerical integration~\cite{Francesco:2019yqt,Bonciani:2019jyb,Czakon:2008zk}, four-dimensional methods~\cite{Gnendiger:2017pys,Heinrich:2020ybq,TorresBobadilla:2020ekr}, contour deformation assisted by neural networks~\cite{Winterhalder:2021ngy}; as well as the achievement of theoretical predictions at fourth order (N$^3$LO) for specific cross-sections~\cite{Camarda:2021ict,Duhr:2020sdp,Currie:2018fgr,Mistlberger:2018etf,Dulat:2017prg}.
All these methodologies may soon be challenged by the theoretical precision required at high-energy colliders.

Despite recent proposals on quantum numerical evaluation of tree-level helicity amplitudes~\cite{Bepari:2020xqi}, it is generally accepted that the perturbative description of hard scattering processes at high energies is beyond the reach of quantum computers, since it would require a prohibitive number of qubits. In this article, we present a proof-of-concept of a quantum algorithm applied to perturbative quantum field theory and demonstrate that the unfolding of certain properties of Feynman loop integrals is fully appropriate and amenable in a quantum computing approach. 

The problem we address is the bootstrapping of the causal representation of multiloop Feynman integrals in the loop-tree duality (LTD) formalism
from the identification of all internal configurations that fulfill causality among the $N=2^n$ potential solutions, where $n$ is the number of internal Feynman propagators. As we will show, this is a satisfiability problem that can be solved with Grover's algorithm~\cite{Grover:1997fa}.
The archetypal situation in which this algorithm is employed consists in finding a single and unique solution among a large unstructured set of $N$ configurations. While a classical algorithm requires testing the satisfiability condition for all cases, i.e. ${\cal O} (N)$ iterations, the quantum algorithm considers all the states in a uniform superposition and tests the satisfiability condition at once. Ultimately, the complexity of the task goes from ${\cal O}(N)$ in the classical case to ${\cal O}(\sqrt{N})$ in the quantum one. This constitutes a big motivation to explore the applicability of such algorithms in the calculation of Feynman diagrams and integrals. Since its introduction in 1996, Grover's algorithm has been generalized~\cite{Brassard:1997gj,Grover:1997ch} and adapted for other applications, such as solving the collision problem~\cite{Brassard:1997aw} or performing partial quantum searches~\cite{2004quant.ph..7122G}. In this article, we introduce a suitable modification of the original Grover's algorithm for querying of multiple solutions~\cite{Boyer:1996zf} to identify all the causal states of a multiloop Feynman diagram. 

From a purely mathematical perspective causal solutions correspond in graph theory to directed acyclic graphs~\cite{squires2020active}, 
which have a broad scope of applications in other sciences, including the characterization of quantum networks~\cite{PhysRevA.80.022339}.
In classical computation, there exist performant algorithms that identify closed
directed loops in connected graphs based on searches on tree representations, such as the well known depth-first search method~\cite{Even20111}.
We apply a different strategy, exploiting the structure of graphs that are relevant in higher-order perturbative calculations, 
in order to ease the identification of causal solutions.

The LTD, initially proposed in Ref.~\cite{Catani:2008xa,Bierenbaum:2010cy,Bierenbaum:2012th}, 
has undergone significant development in recent years~\cite{Buchta:2014dfa,Hernandez-Pinto:2015ysa,Buchta:2015wna,Sborlini:2016gbr,Sborlini:2016hat,Tomboulis:2017rvd,Driencourt-Mangin:2017gop,Jurado:2017xut,Driencourt-Mangin:2019aix,Runkel:2019yrs,Baumeister:2019rmh,Aguilera-Verdugo:2019kbz,Runkel:2019zbm,Capatti:2019ypt,Driencourt-Mangin:2019yhu,Capatti:2019edf,Verdugo:2020kzh,Plenter:2020lop,Aguilera-Verdugo:2020kzc,Ramirez-Uribe:2020hes,snowmass2020,Capatti:2020ytd,Aguilera-Verdugo:2020nrp,Prisco:2020kyb,TorresBobadilla:2021ivx,Sborlini:2021owe,TorresBobadilla:2021dkq,Aguilera-Verdugo:2021nrn}.
One of its most outstanding properties is the existence of a manifestly causal representation, 
which was conjectured for the first time in Ref.~\cite{Verdugo:2020kzh} and further developed in Refs.~\cite{Aguilera-Verdugo:2020kzc,Ramirez-Uribe:2020hes,snowmass2020,Capatti:2020ytd,Aguilera-Verdugo:2020nrp,Sborlini:2021owe,TorresBobadilla:2021ivx}.
A Wolfram Mathematica package, \texttt{Lotty}~\cite{TorresBobadilla:2021dkq}, has recently been released to 
automate calculations in this formalism. The cancellation of noncausal singularities among different contributions 
of the LTD representation of Feynman loop integrals was first observed at one loop in Ref.~\cite{Buchta:2014dfa,Buchta:2015wna} and at higher-orders in Refs.~\cite{Driencourt-Mangin:2019aix,Aguilera-Verdugo:2019kbz,Capatti:2019ypt}.
Noncausal singularities are unavoidable in the Feynman representation of loop integrals, 
although they do not have any physical effect.
Even if they cancel explicitly in LTD among different terms, they lead to significant numerical instabilities. 
Remarkably, noncausal singularities are absent in the causal LTD representation resulting in more stable 
integrands (see e.g. Ref.~\cite{Ramirez-Uribe:2020hes}).
Therefore, the main motivation of this article is to exploit and combine the most recent developments in LTD 
with the exploration of quantum algorithms in perturbative quantum field theory.

The outline of the paper is the following. In Sec. \ref{sec:LTD}, we present a brief introduction to the loop-tree duality (LTD), with special emphasis in the causal structure. In Sec. \ref{sec:Causalflux}, we describe how to efficiently obtain causal configurations by using geometrical arguments. In particular, we motivate the importance of identifying all the configurations with a consistent causal flow of internal momenta, which are equivalent to directed acyclic graphs. Then, we describe the quantum algorithm and its implementation in Sec. \ref{sec:QuantumGrover}. We present explicit examples up to four eloops in Sec. \ref{sec:Application}, where we compare with results already obtained with a classical computation \cite{Verdugo:2020kzh,Aguilera-Verdugo:2020kzc,Ramirez-Uribe:2020hes}. In Sec. \ref{ssec:Counting} we explain the counting of states fulfilling the causality conditions, and how this makes the problem suitable for applying a quantum querying algorithm. Finally, we present our conclusions and comment on possible future research directions in Sec. \ref{sec:Conclusion}.

\section{Causality and the loop-tree duality}
\label{sec:LTD}
Loop integrals and scattering amplitudes in the Feynman representation are defined as integrals in the 
Minkowski space of $L$ loop momenta
\beq
{\cal A}_F^{(L)} = \int_{\ell_1 \ldots \ell_L} {\cal N} (\{\ell_s\}_L, \{p_j\}_P) \prod_{i=1}^n G_F(q_i)~,
\label{eq:AF}
\eeq
where the momentum $q_i$ of each Feynman propagator, $G_F(q_i)$, is a linear combination
of the primitive loop momenta, $\ell_s$ with $s\in \{1,\ldots,L\}$, and external momenta, $p_j$ with $j\in\{1,\ldots, P\}$.
The numerator ${\cal N}$ is determined by the interaction vertices in the given theory and 
the kind of particles that propagate, i.e. scalars, fermions or vector bosons. Its specific form is not relevant 
for the following application. 
The integration measure in dimensional regularization~\cite{Bollini:1972ui,tHooft:1972tcz} is given by
\beq 
\int_{\ell_s} = -\imath \mu^{4-d} \int d^d \ell_s/(2\pi)^d \,,
\eeq
where $d$ is the number of space-time dimensions and $\mu$ is an arbitrary energy scale. 
Rewriting the Feynman propagators in momentum space in the unconventional form 
\beq
G_F(q_i) = \frac{1}{(q_{i,0}-\qon{i})(q_{i,0}+\qon{i})}~,
\eeq
with $\qon{i} = \sqrt{\qb_i^2+m_i^2-\ii}$ (where $\qb_i$ are the spatial components of $q_i$ and $m_i$ is the mass of the propagating particle), one clearly observes that the integrand in \Eq{eq:AF} becomes singular 
when the energy component $q_{i,0}$ takes one of the two values $\pm\qon{i}$. This corresponds to setting 
on shell the Feynman propagator with either positive or negative energy. 
If we always label $q_i$ as  flowing in the same direction, the corresponding time ordered
diagram describes particles propagating forward or backward in time, respectively. 
If we are allowed to modify the momentum flow, the negative energy state represents 
an on-shell particle propagating in the opposite direction as the positive energy one. 
Regardless of our physical interpretation, the two on-shell states of a Feynman propagator are naturally encoded in a qubit and if all the 
propagators get on shell simultaneously there are $N=2^n$ potential singular configurations. 

However, not all potential singular configurations of the integrand lead to physical singularities of the integral. 
The well-known Cutkosky's rules~\cite{Cutkosky:1960sp} provide a simple way to calculate the discontinuities of scattering amplitudes 
that arise when particles in the loop are produced as real particles, requiring that the momentum flow of the 
particles that are set on shell are aligned in certain directions over the threshold cut.
All other singularities are noncausal and should have no physical effect on the integrated expression.
However, they still manifest themselves as singularities of the integrand.

\begin{figure}[ht]
\begin{center}
\includegraphics[scale=1]{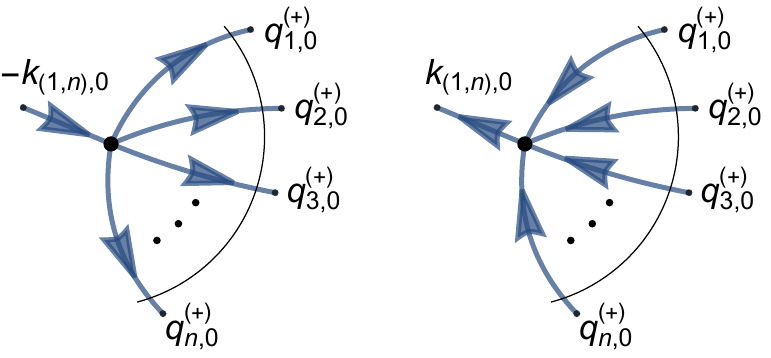}
\end{center}
\caption{Graphical interpretation of the causal configurations encoded by $\lambda^-_{(1,n)}$ (right). 
If the total external momentum is outgoing, $k_{(1,n),0}>0$, a threshold singularity 
arises when all the internal momenta are on shell and their three-momenta flow towards the interaction vertex. The mirror configuration encoded by $\lambda^{+}_{(1,n)}$ (left)
generates a threshold singularity if $k_{(1,n),0}<0$. 
 \label{fig:causaldenominator}}
\end{figure}

In order to have a deeper understanding of the structures leading to the causal singularities, we can exploit the most relevant features of the LTD formalism. The LTD representation of \Eq{eq:AF} is obtained by integrating out one of the components of the 
loop momenta through the Cauchy's residue theorem, then reducing the dimensionality of the integration domain by one 
unit per loop. The integration of the energy component is advantageous because the remaining integration domain, defined by the loop three-momenta, is Euclidean. 
Nevertheless, the LTD theorem is valid in any coordinate system~\cite{Catani:2008xa,Verdugo:2020kzh}.
As a result, Feynman loop integrals or scattering amplitudes are recast as a sum of nested residues, each term representing 
a contribution in which $L$ internal particles have been set on shell in such a way that the loop configuration is open 
to a connected tree. Explicitly, after all the nested residues are summed up, noncausal contributions are analytically cancelled and 
the loop integral in \Eq{eq:AF} takes the causal dual representation
\beq
{\cal A}_D^{(L)} = \int_{\vec \ell_1 \ldots \vec \ell_L} 
\frac{1}{x_n} \sum_{\sigma  \in \Sigma} \frac{{\cal N}_{\sigma(i_1, \ldots, i_{n-L})}}{\lambda_{\sigma(i_1)}^{\sigma(h_1)} \cdots \lambda_{\sigma(i_{n-L})}^{\sigma(h_{n-L})}}
+ (\lambda_p^+ \leftrightarrow \lambda_p^-)~,
\label{eq:AD}
\eeq
with $x_n = \prod_n 2\qon{i}$ and $\int_{\vec \ell_s} = -\mu^{4-d} \int d^{d-1} \ell_s/(2\pi)^{d-1}$ the 
integration measure in the loop three-momentum space.
The Feynman propagators from \Eq{eq:AF} are substituted in \Eq{eq:AD} by causal propagators of the form $1/\lambda_p^\pm$, where
\beq
\lambda_p^\pm = \sum_{i\in p} \qon{i} \pm k_{p,0}~,
\eeq
where $p$ is a partition of the set of on-shell energies, and $k_{p,0}$ is a linear combination of the energy components of the external momenta. 
Causal propagators may appear raised to a power if the Feynman propagators in the original representation 
are raised to a power, for example due to self-energy insertions.
Each $\lambda_p^\pm$ is associated to a kinematic configuration in which the momentum 
flows of all the propagators that belong to the partition $p$ are aligned in the same direction. 
A graphical interpretation is provided in Fig.~\ref{fig:causaldenominator}. Any other configuration cannot be interpreted 
as causal and is absent from \Eq{eq:AD}. Depending on the sign of $k_{p,0}$, either $\lambda_p^+$ or $\lambda_p^-$
becomes singular when all the propagators in $p$ are set on shell. 

The set $\Sigma$ in \Eq{eq:AD} contains all the combinations of causal denominators that
are entangled, i.e. whose momentum flows are compatible with each other and therefore 
represent causal thresholds that can occur simultaneously. Each element in $\Sigma$ fixes the momentum flows of all propagators in specific directions. Conversely, once the momentum flows of all propagators are fixed, the causal representation in \Eq{eq:AD} can be bootstrapped. In the next section, we will explain in more details the geometrical concepts that justify these results, establishing a connection with the formalism presented in Refs.~\cite{TorresBobadilla:2021ivx,Sborlini:2021owe}.

The LTD causal representation has similarities with Cutkosky's rules~\cite{Cutkosky:1960sp} and 
Steinmann's relations~\cite{steinmann,Stapp:1971hh,Cahill:1973qp,Caron-Huot:2016owq,Caron-Huot:2019bsq,Benincasa:2020aoj,Bourjaily:2020wvq,TorresBobadilla:2021ivx}
in that it only exhibits the physical or causal singularities but it is essentially different in that it provides the full integral, and not solely the associated discontinuities.

\section{Geometric interpretation of causal flows}
\label{sec:Causalflux}
Originated from the perturbative expansion of the path integral, multiloop scattering amplitudes are described by Feynman diagrams made of vertices and lines connecting them. Whilst vertices codify interactions among particles, lines are associated to virtual states propagating before/after the interactions take place. These Feynman diagrams might contain closed paths or loops, which symbolise quantum fluctuations involving the emission and subsequent absorption of a virtual particle. As described in the previous section, the number of loops corresponds to the number of free integration variables in Eq. (\ref{eq:AF}).

\begin{figure}[ht]
\begin{center}
\includegraphics[scale=1]{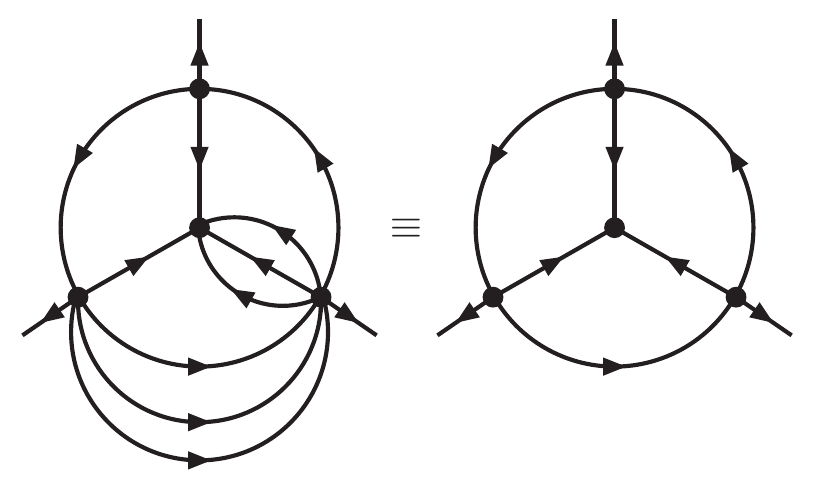}
\end{center}
\caption{Causal equivalence of a multiloop Feynman diagram (left) with a reduced Feynman graph made of edges obtained by merging all propagators connecting a pair of vertices (right).
\label{fig:eloopDEF}}
\end{figure}

However, the dual causal representations presented in Refs. \cite{Verdugo:2020kzh,Aguilera-Verdugo:2020kzc,Ramirez-Uribe:2020hes} can be described by relying on \emph{reduced} Feynman graphs built from vertices and \emph{edges} \cite{TorresBobadilla:2021ivx,Sborlini:2021owe} \footnote{In Ref. \cite{Sborlini:2021owe}, the word \emph{multi-edge} is used instead of edge to avoid confusion with the notation traditionally developed for geometry and graph theory.}. Considering a number of propagators (lines) connecting a pair of interaction vertices, the only possible causal configurations are those in which the momentum flow of all the propagators are aligned in the same direction. As a result, and with the purpose of bootstrapping the causal configurations, a multiloop bunch of propagators can be replaced by a single \emph{edge} representing the common momentum flow \cite{TorresBobadilla:2021ivx,Sborlini:2021owe}, see Fig.~\ref{fig:eloopDEF}. This replacement is further supported by the explicit demonstrations reported in Ref. \cite{Aguilera-Verdugo:2020nrp}.

Once propagators have been collapsed into edges, we can count the number of actual loops in the reduced Feynman graph: these are the so-called \emph{eloops}. We would like to emphasize that the number of eloops is always smaller (or equal) to the number of loops. Whilst the latter counts the number of primitive integration variables, the former refers to a purely graphical and topological property of the reduced Feynman graph.

Following the geometrical description of Feynman diagrams, we introduce a topological classification related to the number of vertices, $V$. In concrete, we define the \emph{order} of a reduced diagram as $k=V-1$, which corresponds to the number of off-shell lines involved in the dual representation. In fact, it can be shown that $k=n-L$, and thus the order of the diagram coincides with the number of causal propagators that are being multiplied in each term of the causal representation in Eq. (\ref{eq:AD}).

At this point, let us comment on the reconstruction of the causal structure and some of the available computational strategies for that purpose. Causal propagators, $1/\lambda_p^{\pm}$, are identified efficiently starting from the connected binary partitions of vertices of the reduced Feynman graph. Once the causal propagators are known, the representation in Eq. (\ref{eq:AD}) can be recovered by identifying all the possible causal compatible combinations of $k$ causal propagators: these are the so-called \emph{causal entangled thresholds}. There are three conditions that determine the allowed
entanglements~\cite{Sborlini:2021owe}:
\begin{enumerate}
 \item The combination of $k$ causal propagators depends on the on-shell energies of all the edges.
 \item The two sets of vertices associated to two causal propagators are disjoint, or one of them is totally included in the other. For instance, if a maximally connected graph (i.e. a graph where all the vertices are connected to each other) is composed by the vertices $V=\{1,2,3,4,5\}$, then $\lambda_1=\{2,3,4\}$ and $\lambda_2=\{1,3,4\}$ cannot be simultaneously entangled since their intersection is not empty. But, $\lambda_1$ and $\lambda_3=\{2,3\}$ are causal-compatible because $\{2,3\} \subset \{2,3,4\}$.
 \item \emph{Causal flow}: The momentum of the edges that crosses a given binary partition of vertices (i.e. each $\lambda_i$ being entangled) must be consistently aligned. Momentum must flow from one partition to a different one.
\end{enumerate}
The strategy to successfully identify the set $\Sigma$ in Eq. (\ref{eq:AD}) consists in following the conditions 1 to 3, in that specific order, as already implemented in Refs. \cite{Sborlini:2021owe,Sborlini:2021nqu}. Remarkably, the third condition can be reinterpreted as the directed graphs associated to the reduced Feynman diagram. Since momenta must exit one partition and enter into a different one, there cannot be closed cycles. This means that condition 3 is equivalent to identifying all possible \emph{directed acyclic graphs} compatible with a given set of causal propagators $\{\lambda_{i_1}^{h_1},\ldots,\lambda_{i_k}^{h_k}\}$. In this way, another reformulation exists
for the causal reconstruction:
\begin{enumerate}
 \item \emph{Causal flow}: Identify all the possible directed acyclic graphs obtained from the original reduced Feynman graph.
 \item Dress each causal configuration with all the possible combinations of entangled causal propagators fulfilling conditions 1-2 of the previous listing.
\end{enumerate}
Both approaches turn out to be equivalent, and this justifies our focus on the detection of causal configurations from the corresponding directed acyclic graphs. However, the identification of directed acyclic graphs is known to be very time-demanding in classical computations (as will be later exposed in Sec. \ref{ssec:Counting}). This motivates the search for alternatives that could provide any possible speed-up. In the following we will explain how to use quantum algorithms for such a purpose. This can be considered as a first step towards a fully quantum approach to the identification of entangled causal thresholds.

\begin{figure}[ht]
\begin{center}
\includegraphics[scale=0.95]{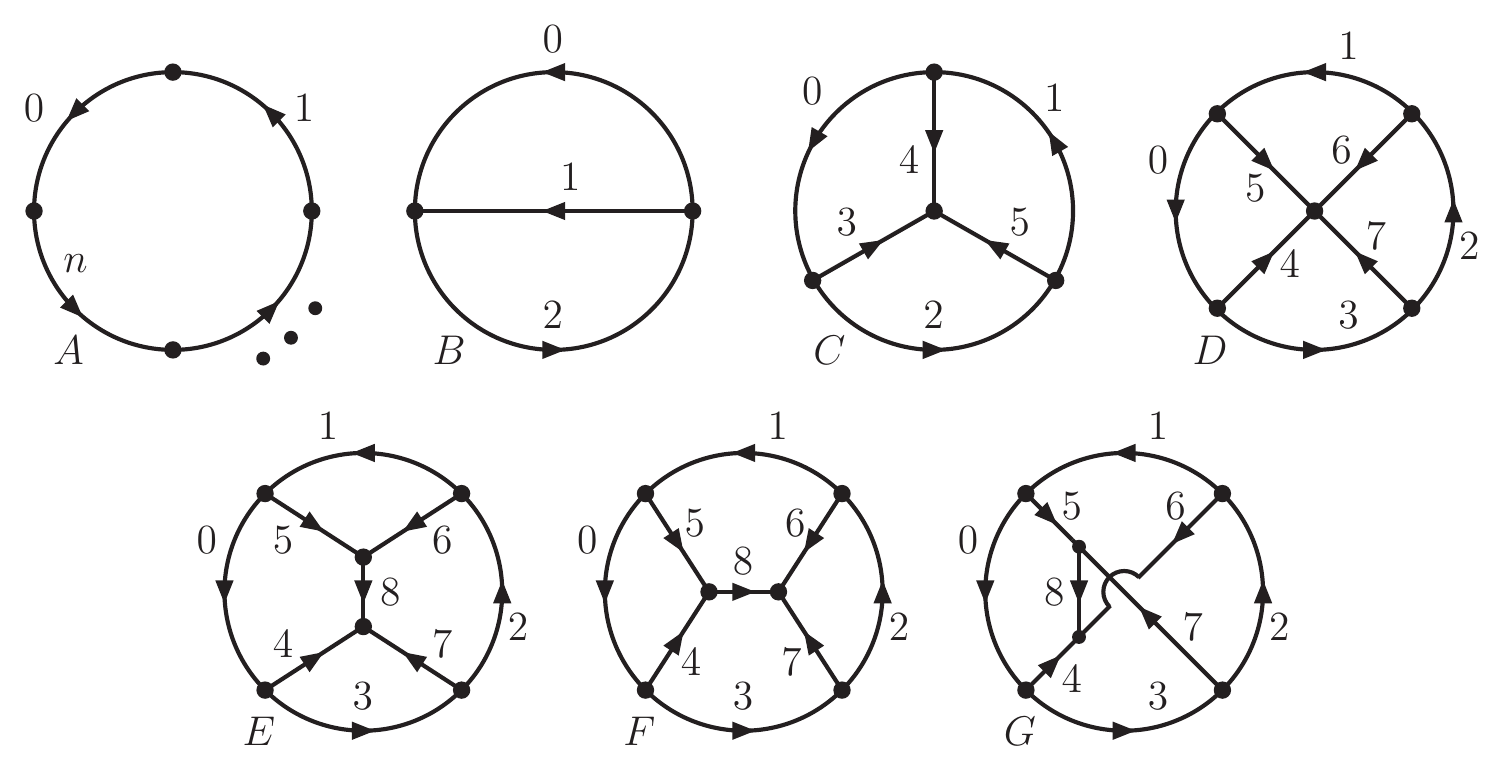}
\end{center}
\caption{Representative multiloop topologies with up to four eloops. The direction of the arrows
corresponds to the $\ket{1}$ states. The vertices may or may not have attached external momenta. Beyond one eloop, each line can be composed of $n_i$ edges that introduce additional vertices. From left to right and top to bottom: one eloop with $n$ vertices, two eloops (MLT), three eloops (N$^2$MLT), four eloops with one four-particle
vertex (N$^3$MLT), and four eloops with trivalent interactions (N$^4$MLT), $t$-, $s$- and $u$-channels. 
\label{fig:multilooptopologies}}
\end{figure}

In Fig.~\ref{fig:multilooptopologies}, we show the representative multiloop topologies that we have considered in this work. We follow the classification scheme introduced in Refs.~\cite{Aguilera-Verdugo:2020kzc,Ramirez-Uribe:2020hes}, where loop diagrams are ranked according to the number of sets of propagators that depend on different linear combinations of the loop momenta, starting from the maximal loop topology (MLT) with $L+1$ sets, to N$^k$MLT with $L+1+k$ sets. An extended classification has been introduced in Ref.~\cite{TorresBobadilla:2021ivx} that considers all the vertices connected to each other.

\section{Quantum algorithm for causal querying}
\label{sec:QuantumGrover}
Following the standard Grover's querying algorithm~\cite{Grover:1997fa} over unstructured databases, we start from 
a uniform superposition of $N=2^n$ states
\beq
\ket{q}= \frac{1}{\sqrt{N}} \sum_{x=0}^{N-1} \ket{x}~, 
\eeq
which can also be seen as the superposition of one winning state $\ket{w}$,
encoding all the causal solutions in a uniform superposition, and the orthogonal state $\ket{q_\perp}$,
that collects the noncausal states
\beq
\ket{q} =  \cos \theta \, \ket{q_\perp} + \sin\theta \, \ket{w}~.
\eeq
The mixing angle is given by $\theta =  \arcsin \sqrt{r/N}$, 
where $r$ is the number of causal solutions, 
and the winning and orthogonal uniform superpositions are given by 
\beq
\ket{w}= \frac{1}{\sqrt{r}} \sum_{x\in w} \ket{x}~, \qquad 
\ket{q_\perp}= \frac{1}{\sqrt{N-r}} \sum_{x\notin w} \ket{x}~.
\eeq
The algorithm requires two operators, the oracle operator
\beq
U_w = \id - 2\ket{w} \bra{w}~, 
\eeq
that flips the state $\ket{x}$ if $x\in w$, $U_w \ket{x} = - \ket{x}$, and leaves it unchanged otherwise,
$U_w \ket{x} = \ket{x}$ if $x\notin w$, and the Grover's diffusion operator
\beq
U_q = 2 \ket{q} \bra{q} - \id~,
\eeq
that performs a reflection around the initial state $\ket{q}$. The iterative application of both operators 
$t$ times leads to 
\beq
(U_q U_w )^t \ket{q} = \cos \theta_t \, \ket{q_\perp } +  \sin \theta_t \, \ket{w}~,
\label{eqAmplifica}
\eeq
where $\theta_t = (2t +1) \, \theta$. The goal is then to reach a final state such that 
the probability of each of the components in the orthogonal state is 
much smaller than the probability of each of the causal solutions by choosing $\theta_t$ accordingly:
\beq
\frac{\cos^2 \theta_t}{N-r} \ll \frac{\sin^2 \theta_t}{r}~.
\label{eq:componentprob}
\eeq
This goal is achieved when $\sin^2 \theta_t \sim 1$.

Grover's standard algorithm works well if $\theta \leq \pi/6$, namely $r \leq N/4$,
but does not provide the desired amplitude amplification of the winning states for larger angles. 
For example, if $\theta = \pi/3$ the first iteration leads to $\theta_1=\pi$ 
which in fact suppresses the projection onto the set of solutions, while for $\theta=\pi/4$ or $r=N/2$ no matter how many iterations are 
enforced the probabilities of the initial states remain unchanged. One of the strategies that we apply, which is also valid for other problems where 
the number of solutions is larger than $N/4$, is to enlarge the total number of 
states without increasing the number of solutions by introducing ancillary qubits in the register
that encodes the edges of the loop diagram \footnote{This strategy has been previously discussed in Ref. \cite{Nielsen2000}.}. 
In general, the maximum number of ancillary qubits needed is two, 
as this increases the number of total states by a factor of $4$. Furthermore, for Feynman loop diagrams we will take advantage of the fact that given a causal solution (directed acyclic configuration), the mirror state in which all internal momentum flows are reversed is also a causal solution. 
Therefore, we will single out one of the edges and consider that only one of its states contributes to the winning set, 
while the mirror states are directly deduced from the selected causal solutions.
As a result, the complete set of causal solutions can be determined with the help of at most one ancillary qubit.

Three registers are needed for the implementation of the quantum algorithm, 
together with another qubit that is used as marker by the Grover's oracle.  
The first register, whose qubits are labelled $q_i$, encode the states of the edges.
The qubit $q_i$ is in the state $\ket{1}$ if the momentum flow of the corresponding edge 
is oriented in the direction of the initial assignment and in $\ket{0}$ if it is in the opposite direction (see Fig.~\ref{fig:multilooptopologies}). 
In any case, the final physical result is independent of the initial assignment, 
being used only as a reference.

The second register, named $c_{ij}$, stores the Boolean clauses 
that probe whether or not two qubits representing two adjacent edges 
are in the same state (whether or not are oriented in the same direction). 
These binary clauses are defined as
\bea
&&c_{ij} \equiv (q_i = q_j)~, \nn \\
&&\bar c_{ij} \equiv (q_i \ne q_j)~, \qquad i,j \in\{0, \ldots, n-1\}~.
\eea
The third register, $a_k(\{c_{ij}\},\{\bar c_{ij}\})$, encodes the loop clauses that 
probe if all the qubits (edges) in each of the eloops that are part of the diagram  
form a cyclic circuit. 

The causal quantum algorithm is implemented as follows. 
The initial uniform superposition is obtained by applying Hadamard gates to each of the qubits 
in the $q$-register, $\ket{q} = H^{\otimes n} \ket{0}$, while the qubit which is used as Grover's marker
is initialized to
\beq
\ket{out_0} =   \frac{\ket{0} - \ket{1}}{\sqrt{2}} \equiv \ket{-} \, ,
\eeq
which corresponds to a Bell state in the basis $\{\ket{0}, \ket{1}\}$. The other registers,
$\ket{c}$ and $\ket{a}$, used to store the binary and eloop clauses are initialized to $\ket{0}$.
Each binary clause $\bar c_{ij}$ requires two CNOT gates operating between two qubits in the $\ket{q}$
register and one qubit in the $\ket{c}$ register. An extra XNOT gate acting on the corresponding qubit in 
$\ket{c}$ is needed to implement a $c_{ij}$ binary clause. 

The oracle is defined as
\beq
U_w \ket{q} \ket{c} \ket{a} \ket{out_0} = (-1)^{f(a,q)} \, \ket{q} \ket{c} \ket{a} \ket{out_0}~.
\eeq
Therefore, if all the causal conditions are satisfied, $f(a,q) = 1$, the corresponding states are marked;
otherwise, if $f(a,q) = 0$, they are left unchanged. After the marking, the $\ket{c}$ and $\ket{a}$ registers 
are rotated back to $\ket{0}$ by applying the oracle operations in inverse order.
Then, the diffuser $U_q$ is applied to the register $\ket{q}$.
We use the diffuser described in the IBM Qiskit website \footnote{\texttt{http://qiskit.org/}}.

\begin{table}
\begin{center}
    \begin{tabular}{ccccc} \hline \hline 
eloops (edges per set)& $\ket{q}$ & $\ket{c}$ & $\ket{a}$ & Total\\ \hline
one ($n$) & $n+1$ & $n-1$ & $1$ & $2 n+2$\\
two ($n_0, n_1, n_2$) & $n$ & $n$ &$3$ & $2 n + 4$\\
three ($n_0, \ldots, n_5$) & $n$ & $n+ (2$ to $3)$ & $4$ to $7$ & $2 n + (7$ to $11)$\\
four ($n_0, \ldots, n_7$) & $n$ & $n + (3$ to $6)$ & $5$ to $13$ & $2 n + (9$ to $20)$\\
four ($n_0, \ldots, n_8^{(t,s)}$) & $n$ & $n + (4$ to $7)$ & $5$ to $13$ & $2 n + (10$ to $21)$\\
four ($n_0, \ldots, n_8^{(u)}$) & $n$ & $n + (5$ to $8)$ & $9$ to $13$ & $2 n + (15$ to $22)$\\
\hline \hline 
\end{tabular}
\end{center}
\caption{Number of qubits in each of the three main registers. The total number of qubits includes 
the ancillary qubit  which is initialized to $\ket{-}$ to implement Grover's oracle. Measurements are made on $n=\sum n_i$ 
classical bits. \label{tb:qubits}} 
\end{table}

The upper and lower limit in the number of qubits needed to analyze loop topologies of up to four eloops is 
summarized in Tab.~\ref{tb:qubits}. The final number of qubits depends on the 
internal configuration of the loop diagram. The lower limit is achieved if $n_i =1$ for all the sets, the upper 
limit is saturated for $n_i\ge 2$.
Specific details on the implementation of the quantum 
algorithm and causal clauses are provided in the next section. 
We use two different simulators: \emph{IBM Quantum} provided by the open source Qiskit framework; and \emph{Quantum Testbed} (QUTE) \cite{alonso_raul_2021_5561050}
, a high performance quantum simulator developed and maintained by Fundaci\'on Centro Tecnol\'ogico de la Informaci\'on y la Comunicaci\'on (CTIC) \footnote{\texttt{http://qute.ctic.es/}}.

The output of the Grover's algorithm described above is a quantum state that is predominantly a superposition of the whole set of causal solutions, with a small contribution from orthogonal states. After a measurement, a single configuration is determined and the superposition is lost. If one requires knowing all solutions and not just a single one, the original output of Grover's algorithm has then to be prepared and measured a certain number of times, also called shots, large enough in order to scan over all causal solutions, and to distinguish them from the less probable noncausal states. 
The final result is represented by frequency histograms and is affected by the statistical fluctuations that are inherent to the measurements of a quantum system. Our approach is based on Grover's search algorithm and, as such, has a similar quantum depth compared to the original implementation and thus a well-known noisy performance on a real present device~\cite{9151202,PhysRevA.102.042609,QuantumInfProcess20}. Given the quantum depth of the algorithm and the resulting difficulties in introducing a reliable error mitigation strategy, we will only consider error-free statistical uncertainties in quantum simulators. Nevertheless, for the sake of benchmarking, we will present a simulation on a real device for the less complex multiloop topology we have analyzed.

We estimate that the number of shots required to distinguish causal from noncausal configurations with a statistical significance of $\Delta \sigma$ standard deviations in a quantum simulator is given by 
\beqn
N_{\rm shots} \approx r \left(\Delta \sigma\right)^2 \left(1 + {\cal O}(\cos^2(\theta_t)) \right) \, ,
\label{eq:shots}
\eeqn
assuming that an efficient amplification of the causal states is achieved, i.e. $\cos(\theta_t) \sim 0$.

For the identification of causal configurations of the multiloop topologies shown in Fig.~\ref{fig:multilooptopologies}, for which the number of solutions is of the order of $1/4$ of the total number of states, the quantum advantage over classical algorithms is suppressed by the number of required measurements given by \Eq{eq:shots}. However, as we will explain in Sec. \ref{ssec:Counting}, the number of states fulfilling causal-compatible conditions for increasingly complex topologies is much smaller than the total combinations of thresholds. Thus, for very complex topologies which are less affordable with a classical computation, we turn back to the original quantum speed-up provided by Grover's algorithm. In the following, we will consider $\Delta \sigma \gtrsim 3$, which provides a sufficiently safe discriminant yield with a minimal number of shots.

\section{Benchmark multiloop topologies}
\label{sec:Application}
After introducing the quantum algorithm that identifies the causal configurations of multiloop Feynman diagrams in Sec. \ref{sec:QuantumGrover} and explaining the connection between acyclic graphs and causality in Sec. \ref{sec:Causalflux}, we present here concrete examples. We consider several topological families of up to four eloops, discussing in each case the explicit implementation of the Boolean clauses and explaining the results obtained.

\subsection{One eloop}
\label{ssec:1eloop}
The one-eloop topology consists of $n$ vertices connected with $n$ edges along a one loop circuit (see Fig.~\ref{fig:multilooptopologies}A). Each vertex has an external particle attached to it, although it is also possible to have vertices without attached external momenta that are the result of collapsing, e.g., a self-energy insertion into a single edge as explained in Sec.~\ref{sec:Causalflux}.

\begin{figure}[t]
\begin{center}
\includegraphics[trim={5cm 1.8cm 2.5cm 1.5cm}, width=420px]{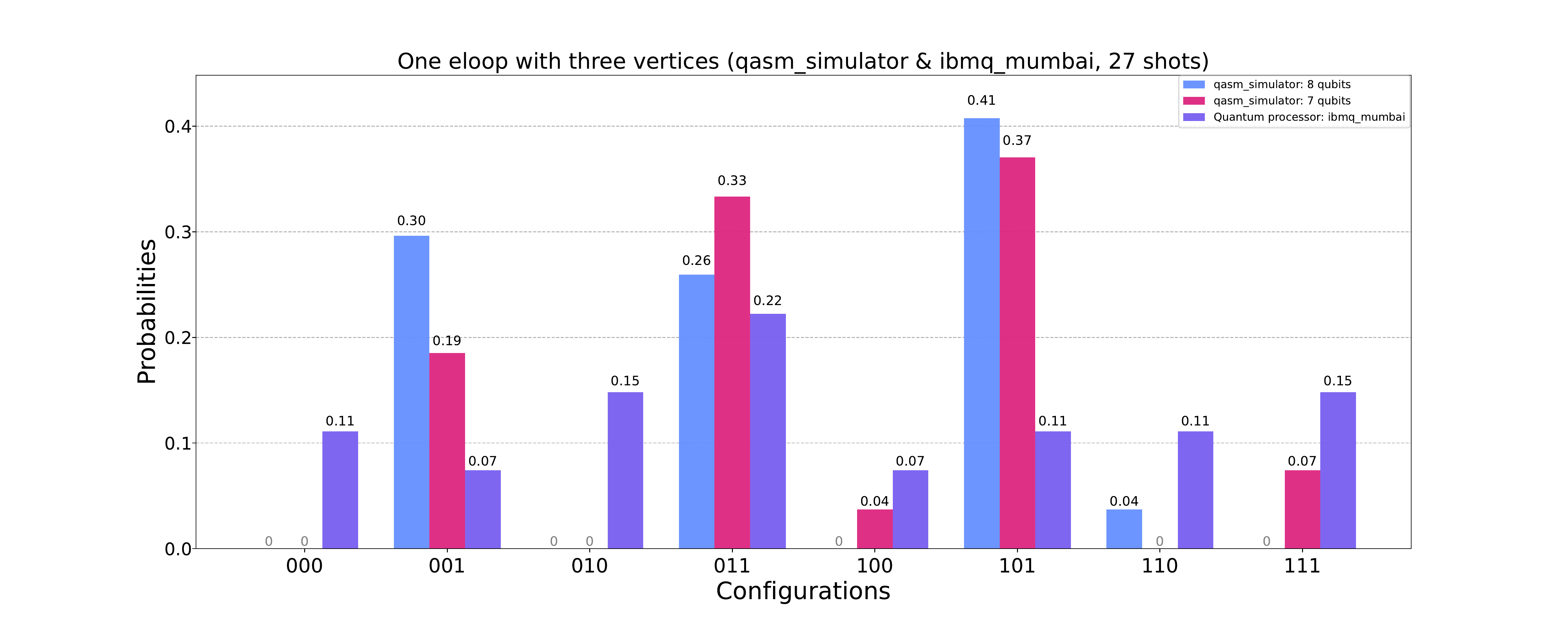}
\end{center}
\caption{Probability distribution of causal and noncausal configurations 
obtained with (blue) and without (purple) 
an ancillary qubit for a one-eloop three-vertex topology.
Results are presented with both the IBM quantum simulator and the real quantum device.
\label{fig:prob3edge}}
\end{figure}


\begin{figure}[t]
\begin{center}
\includegraphics[width=360px]{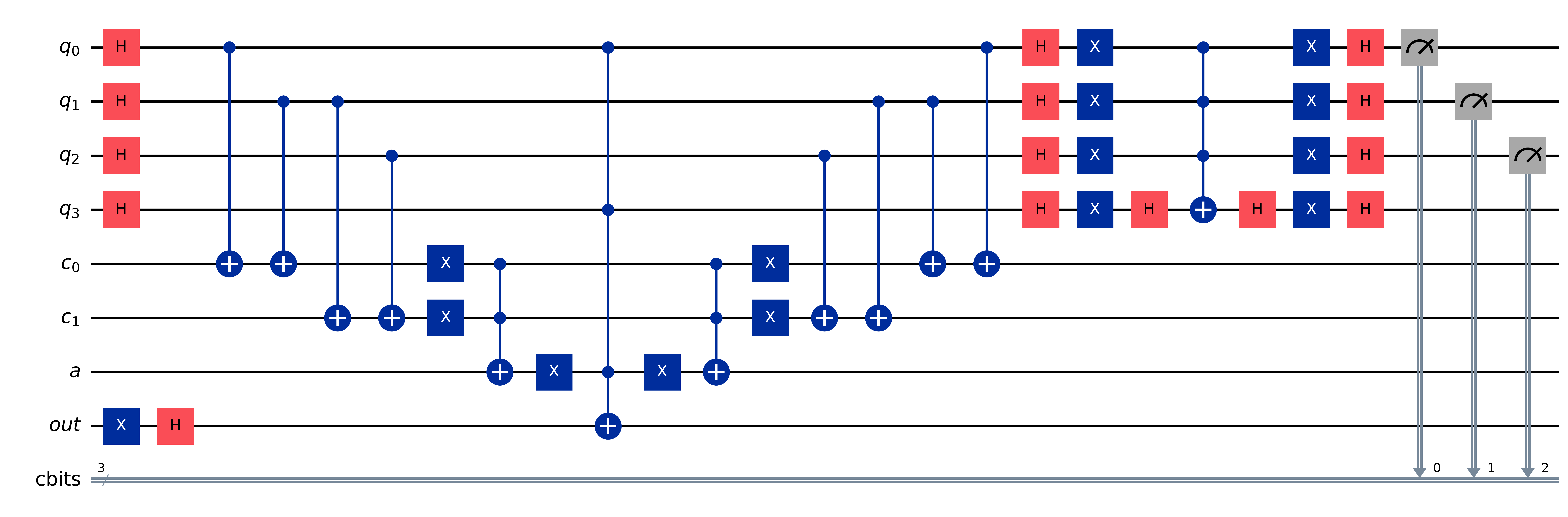} \\
\includegraphics[width=360px]{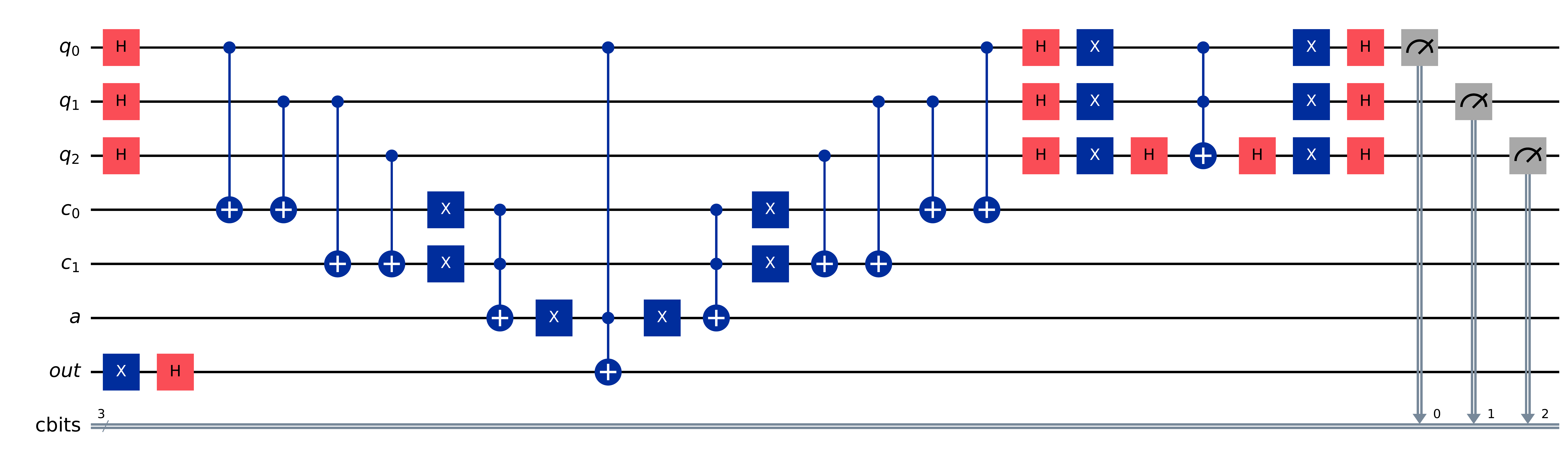}
\end{center}
\caption{Quantum circuits used to bootstrap the causal configuration of a three-vertex multiloop Feynman diagram. Implementation with (up) and without (down) an ancillary qubit. \label{fig:qc3edge}}
\end{figure}

\begin{figure}[t]
\begin{center}
\includegraphics[scale=0.95]{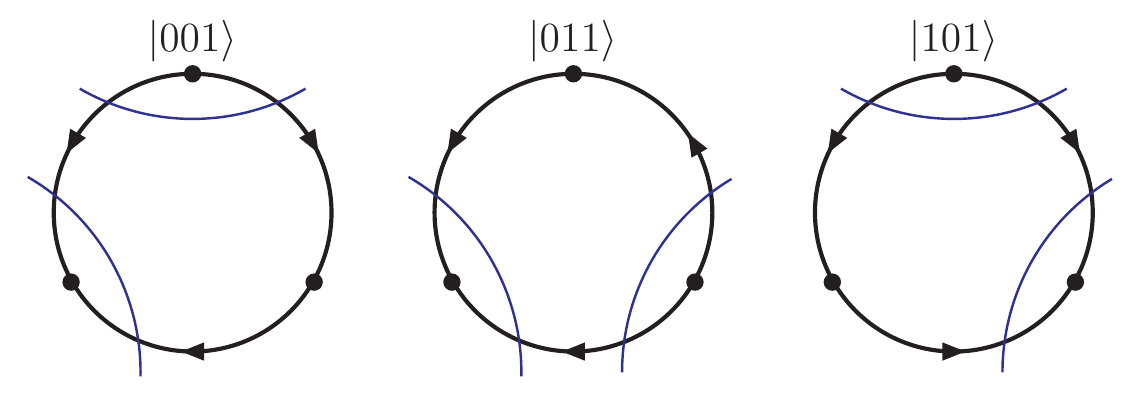}
\end{center}
\caption{Causal bootstrapping of the one-eloop three-vertex topology. 
Configurations with all internal momentum flows reversed (not shown) are also causal. 
\label{fig:causal3edge}}
\end{figure}

We need to check $n-1$ binary clauses, and there is one Boolean 
condition that has to be fulfilled
\beq
a_0(\{c_{ij}\}) \equiv \neg \left( c_{01} \wedge c_{12} \wedge \cdots \wedge c_{n-2,n-1} \right)~.
\eeq
The qubit $a_0$ is set to one if not all the edges are oriented in the same direction.
This condition is implemented by imposing a multicontrolled Toffoli gate followed by a Pauli-X gate.
We know, however, that this condition is fulfilled for $N-2$ states at one eloop.
Therefore, the initial Grover's angle tends to $\left. \arcsin \left( \sqrt{(N-2)/N} \right) \right|_{n\to \infty} = \pi/2$.
In order to achieve the suppression of the orthogonal states, we introduce one ancillary qubit, $q_n$, 
and select one of the states of one of the qubits representing one of the edges. 
The required Boolean marker is given by 
\beq
f^{(1)}(a,q) = a_0 \wedge q_0 \wedge q_{n}~,
\eeq
which is also implemented through a multicontrolled Toffoli gate. 

The ratio of probabilities of measuring a winning state versus an orthogonal state is enhanced by adding the ancillary qubit.
Alternatively, we can still rely on the original Grover's algorithm when the number of noncausal configurations is small, by swapping the definition of winning and orthogonal states. However, the ancillary qubit is absolutely necessary when the number of winning solutions is ${\cal O}(N/2)$.


The output of the algorithm for a three-vertex multiloop topology is illustrated in  Fig.~\ref{fig:prob3edge}, 
where we extract and compare the selection of causal states with and without the ancillary qubit. The corresponding quantum circuits are represented in Fig.~\ref{fig:qc3edge}. The ancillary qubit is set 
in superposition with the other qubits but is not measured because this information is irrelevant. 
Note that in the Qiskit convention qubits are ordered in such a way that the last qubit
appears on the left-most side of the register~$\ket{q}$.

Fig.~\ref{fig:causal3edge} shows the corresponding directed acyclic configurations 
and the bootstrapped causal interpretation in terms of causal thresholds. 
Once the direction of the edges is fixed by the quantum algorithm,
the causal thresholds are determined by considering all the possible on-shell cuts with aligned edges that are compatible or entangled with each other. This information can be translated directly 
into the LTD causal representation in Eq.~(\ref{eq:AD}); the on-shell energies $\qon{i}$ that contribute 
to a given causal denominator, $\lambda_p^\pm$, are those related through the same threshold.

The quantum depth of the circuit estimated by Qiskit is $25$ in the simulator, while it amounts to ${\cal O}(200)$ with the ancillary qubit, and ${\cal O}(150)$ without the ancillary qubit in a real device where not all the qubits are connected to each other. The circuit depth is therefore too large to provide a good result in a present real device, as illustrated in Fig.~\ref{fig:prob3edge}. We will focus hereafter on the results obtained by quantum simulators. They are in full agreement with the expectations.

\begin{figure}[ht]
\begin{center}
\includegraphics[scale=0.99]{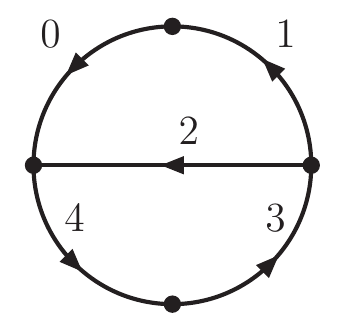}
\end{center}
\caption{Two-eloop five-edge topology.}
\label{fig:fiveedgetwoloop} 
\end{figure}

\begin{figure}[t]
\begin{center}
\includegraphics[width=300px]{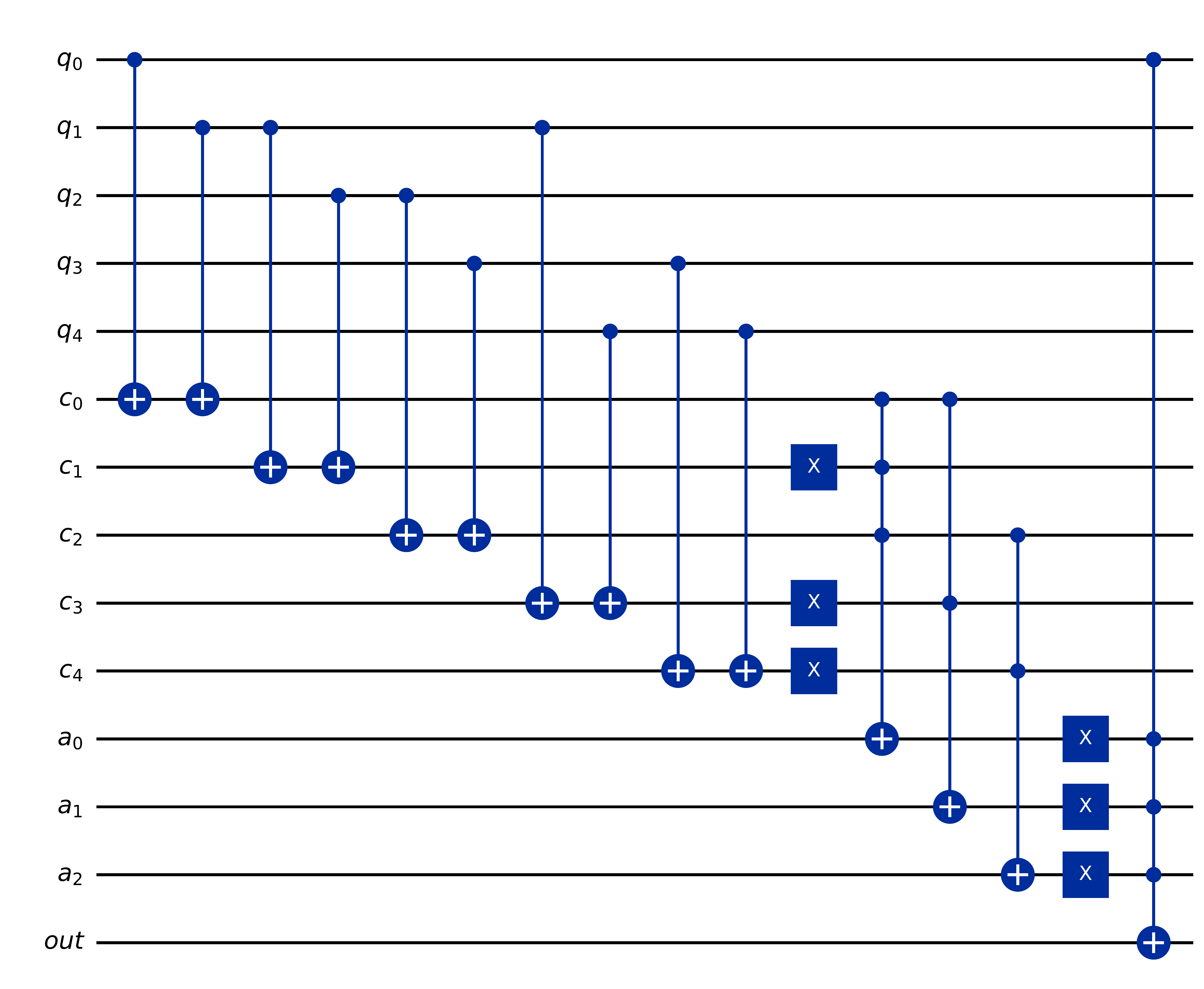} \\
\includegraphics[width=420px]{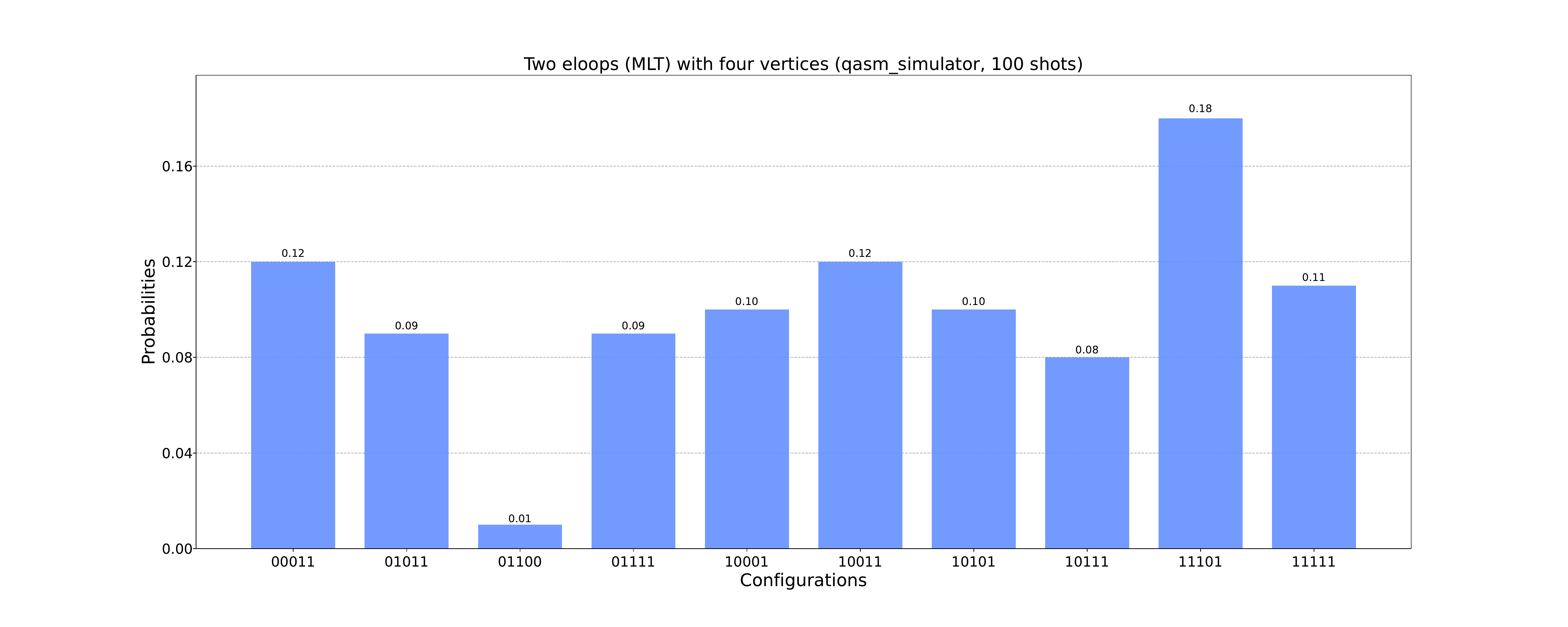}
\end{center}
\caption{Oracle of the quantum circuit (up, omitting the reflection of the quantum gates) and probability distribution of causal and noncausal configurations (down) 
for a two-eloop topology (MLT) with $n_0=n_2=2$ and $n_1=1$ edges.
\label{fig:mlt5}}
\end{figure}

\begin{figure}[t]
\begin{center}
\includegraphics[scale=0.4]{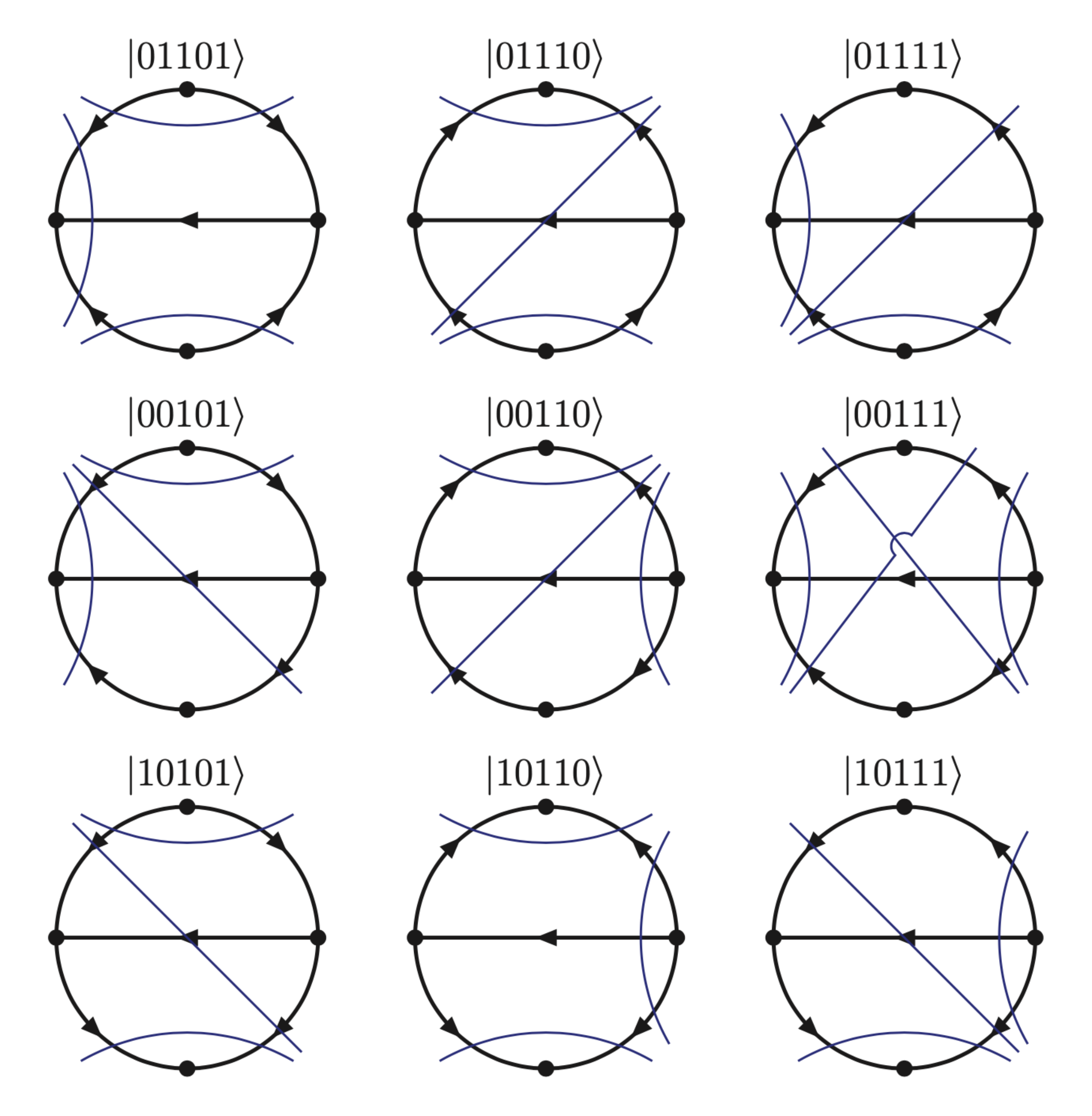}
\end{center}
\caption{Causal bootstrapping of the two-eloop five-edge topology. Configurations with all internal momentum flows reversed (not shown) are also causal. 
\label{fig:causalmlt5}}
\end{figure}

\subsection{Two eloops}
\label{ssec:2eloops}
We now analyze multiloop topologies with two eloops (see Fig.~\ref{fig:multilooptopologies}B).
These topologies are characterized by three sets of edges with $n_0$, $n_1$
and $n_2$ edges in each set and two common vertices. The first non-trivial 
configuration requires that at least two of the sets contain two or more edges. 
If $n_0=n_1=n_2=1$, we have a multibanana or MLT configuration with $L+1$ propagators which is equivalent to one edge,
while the NMLT configuration with $L+2$ sets of propagators, or $n_0=2$ and $n_1=n_2=1$, is equivalent to the one-eloop three-vertex topology already analyzed in Sec.~\ref{ssec:1eloop}
because propagators in the sets 1 and 2 can be merged into a single edge.
We consider the five-edge topology depicted in Fig.~\ref{fig:fiveedgetwoloop} as the first non at two ops.
The diagram is composed by three subloops, and therefore requires 
to test three combinations of binary clauses
\bea
&& a_0 = \neg \left( c_{01} \wedge c_{13}\wedge c_{34}\right)~, \nn \\
&& a_1 = \neg \left( c_{01} \wedge \bar c_{12} \right)~,\nn \\
&& a_2 = \neg \left( c_{23} \wedge c_{34} \right)~.
\label{eq:astwoeloops}
\eea
We know from a classical computation~\cite{Aguilera-Verdugo:2020kzc} that the number of causal solutions over 
the total number of states is $18/32 \sim 1/2$.
Therefore, it is sufficient to fix the state of one of the edges to reduce the number of states queried to less than $1/4$, 
while the ancillary $q_n$-qubit is not necessary.
We select $q_2$ as the qubit whose state is fixed, and check the Boolean condition
\beq
f^{(2)}(a,q) = (a_0 \wedge a_1 \wedge a_2) \wedge q_2~.
\eeq

The oracle of the quantum circuit and its output in the IBM's Qiskit simulator are shown in Fig.~\ref{fig:mlt5}, and the causal interpretation 
is provided in Fig.~\ref{fig:causalmlt5}. The number of states selected in Fig.~\ref{fig:mlt5} is $9$, corresponding to $18$ causal states 
when considering the mirror configurations obtained by inverting the momentum flows, 
and in full agreement with the classical calculation.

The generalization to an arbitrary number of edges
requires to check first if all the edges in each set are aligned. We define
\beq
b_s = \underset{i_s \in s}{\wedge} c_{i_s, i_s+1}~, \qquad s \in \{0,1,2\}~.
\eeq
The number of subloops is always three, and so the number of conditions that generalize \Eq{eq:astwoeloops}
%
\bea
&& a_0 = \neg \left( b_0 \wedge c_{0_0 (n_2-1)} \wedge b_2 \right)~, \nn \\ 
&& a_1 = \neg \left( b_0 \wedge \bar c_{(n_0-1) (n_1-1)} \wedge b_1 \right)~, \nn \\ 
&& a_2 = \neg \left( b_1 \wedge c_{(n_1-1) 0_2} \wedge b_2 \right)~, 
\eea

\noindent
where $0_s$ represents the first edge of the set $s$, and $(n_s-1)$ is the last one. 
The total number of qubits required to encode these configurations is summarized in Tab.~\ref{tb:qubits}.
With 32 qubits as the upper limit in the IBM Qiskit simulator, one can consider any two-eloop topology
with $\sum n_i \le 14$ distributed in three sets.

\begin{figure}[t]
\begin{center}
\includegraphics[width=300px]{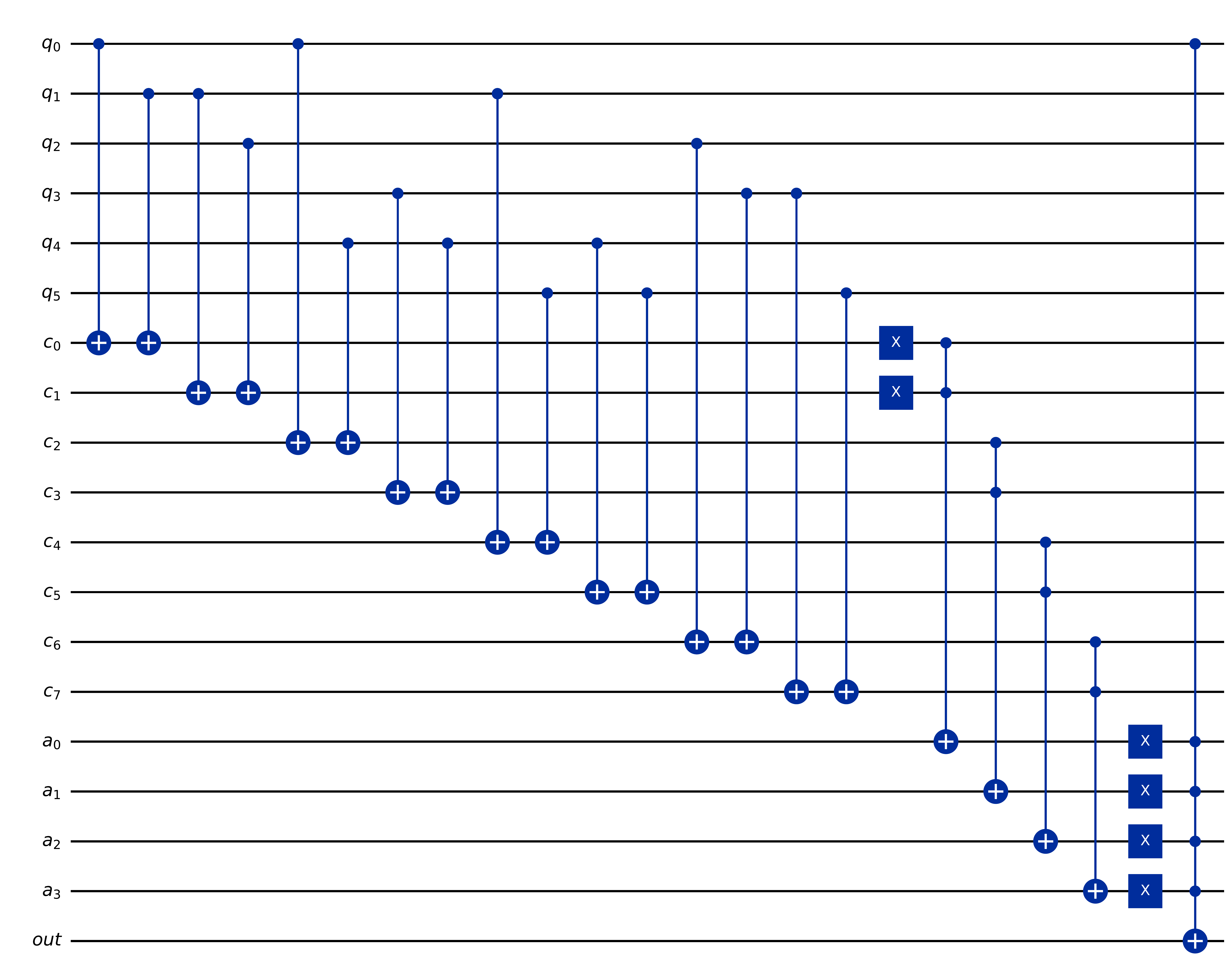} \\
\includegraphics[width=420px, height=200px]{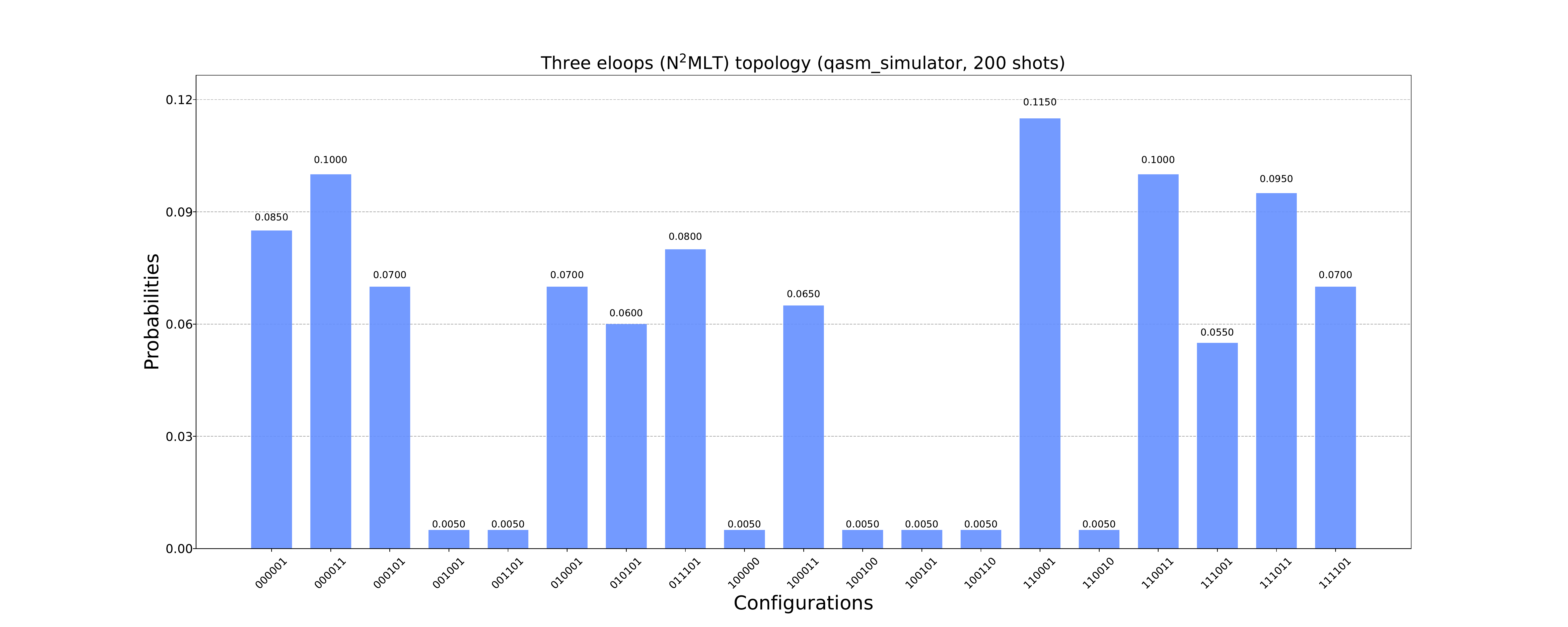}
\end{center}
\caption{Oracle of the quantum circuit (up, omitting the reflection of the quantum gates) and probability distribution of causal and noncausal configurations (down) for a three-eloop topology (Mercedes topology or N$^2$MLT with $n_i=1$).
\label{fig:n2mlt}}
\end{figure}

\begin{figure}[t]
\begin{center}
\includegraphics[scale=0.95]{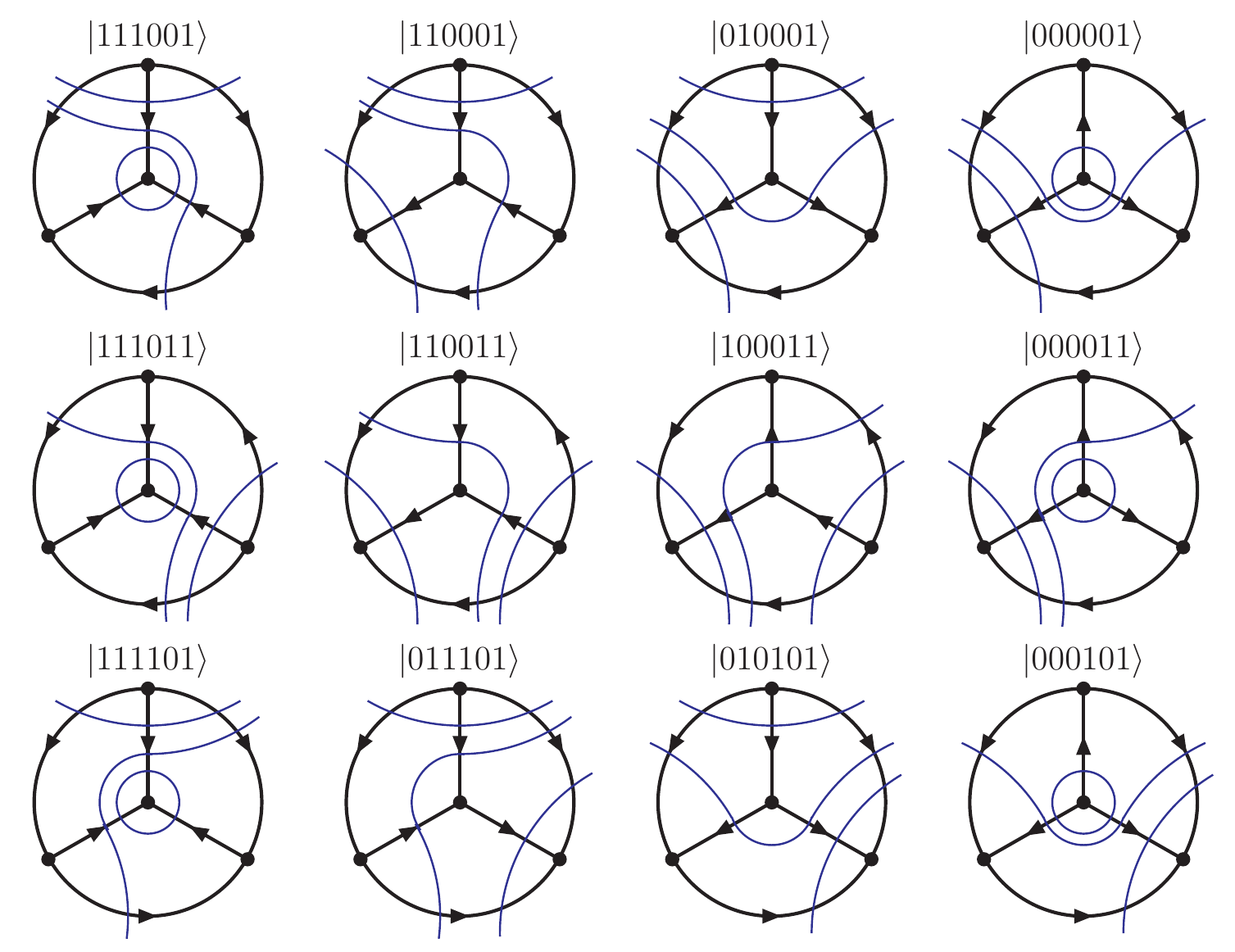}
\end{center}
\caption{Causal bootstrapping of the three-eloop topology (N$^2$MLT). Configurations with all internal momentum flows reversed (not shown) are also causal. 
\label{fig:n2mlt_th}}
\end{figure}

\subsection{Three eloops}
\label{ssec:3eloops}
The N$^2$MLT multiloop topology (see Fig.~\ref{fig:multilooptopologies}C) is characterized by four vertices connected 
through six sets of edges, and $n_i$ edges in each set, $i \in \{0,\ldots, 5\}$.
It appears for the first time at three loops. The algorithm for the multiloop topology 
with $n_i=1$ requires to test the following loop clauses
\bea
&& a_0 = \neg \left( c_{01} \wedge c_{12}\right)~, \nn \\
&& a_1 = \neg \left( \bar c_{04} \wedge \bar c_{34} \right)~,\nn \\
&& a_2 = \neg \left( \bar c_{15} \wedge \bar c_{45} \right)~, \nn \\
&& a_3 = \neg \left( \bar c_{23} \wedge \bar c_{35} \right)~.
\label{eq:a3eloops}
\eea
It is worth noticing that the loop clauses can be implemented in several ways. 
For example the following expressions are equivalent
\beq
\left( \bar c_{04} \wedge \bar c_{34} \right) = \left( c_{03} \wedge \bar c_{34} \right)~.
\label{eq:equiv}
\eeq
However, the expression on the l.h.s. of \Eq{eq:equiv} requires one NOT gate less than the one on the r.h.s., 
so it is preferable. It is also worth mentioning that 
testing loop clauses involving four edges, such as
\beq
\neg ( \bar c_{35} \wedge  \bar c_{15} \wedge c_{01})~,
\label{eq:fouredge}
\eeq
is not necessary because four-edge loops enclose one qubit that in any of its states would 
create a cyclic three-edge loop if the other four edges are oriented in the same direction.
The final Boolean condition is 
\beq
f^{(3)}(a,q) = (a_0 \wedge \ldots \wedge a_3) \wedge q_0~.
\eeq
The oracle of the quantum circuit and probability distribution are shown in Fig.~\ref{fig:n2mlt}, 
and the causal interpretation is given in Fig.~\ref{fig:n2mlt_th}.
The number of causal configurations is $24$ out of $64$ potential configurations.

For configurations with an arbitrary number of edges the loop clauses in 
\Eq{eq:a3eloops} are substituted by 
\bea
&& a_0 = \neg \left( b_0 \wedge c_{(n_0-1) 0_1} \wedge b_1 \wedge c_{(n_1-1) 0_2} \wedge b_2 \right)~, \nn \\
&& a_1 = \neg \left( b_0 \wedge \bar c_{(n_0-1) (n_4-1)} \wedge b_4 \wedge \bar c_{0_4 0_3} \wedge b_3 \right)~,\nn \\
&& a_2 = \neg \left( b_1 \wedge \bar c_{(n_1-1) (n_5-1)} \wedge b_5 \wedge \bar c_{0_5 0_4} \wedge b_4 \right)~, \nn \\
&& a_3 = \neg \left( b_2 \wedge \bar c_{(n_2-1) (n_3-1)} \wedge b_3  \wedge \bar c_{0_3 0_5} \wedge b_5 \right)~.
\label{eq:a3eloopsARBITRARY}
\eea

This is the minimal number of loop clauses at three eloops. 
For three-eloop configurations with several edges in each set an extra binary clause ($c_{(n_2-1) 0_0}$) 
and up to three loop clauses may be needed to test cycles over four edge sets.  
These clauses are
\bea
&& a_4 = \neg \left( b_0 \wedge c_{(n_0-1) 0_1} \wedge b_1 \wedge \bar c_{(n_1-1) (n_5-1)} \wedge b_5 \wedge \bar c_{0_3 0_5}\wedge b_3 \right)~, \nn \\
&& a_5 = \neg \left( b_1 \wedge c_{(n_1-1) 0_2} \wedge b_2 \wedge \bar c_{(n_2-1) (n_3-1)} \wedge b_3 \wedge \bar c_{0_4 0_3}\wedge b_4 \right)~, \nn \\
&& a_6 = \neg \left( b_2 \wedge c_{(n_2-1) 0_0} \wedge b_0 \wedge \bar c_{(n_0-1) (n_4-1)} \wedge b_4 \wedge \bar c_{0_5 0_4}\wedge b_5 \right)~.
\eea
The number of qubits  reaches the upper limit reflected in Tab.~\ref{tb:qubits}
for $n_i\ge 2$.

\begin{figure}[tbh]
\begin{center}
\includegraphics[scale=0.95]{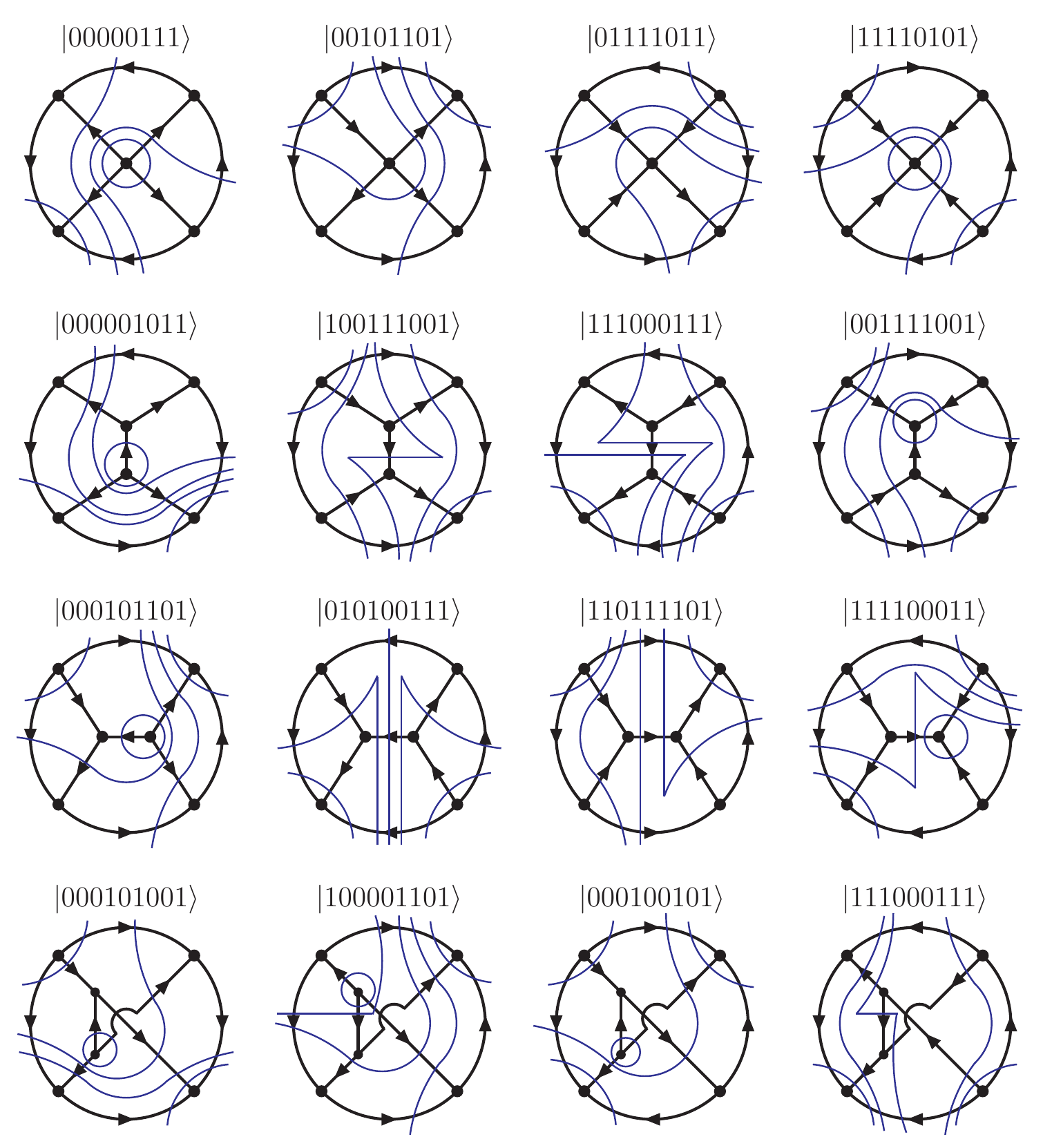}
\end{center}
\caption{Representative bootstrapped diagrams at four eloops, from the quantum algorithm output in Fig.~\ref{fig:prob4eloops}.
\label{fig:4eloops_th}}
\end{figure}

		\begin{figure}[ht]
			\centering
			\begin{minipage}{0.49\textwidth}   
				\centering
				\includegraphics[width=\textwidth]{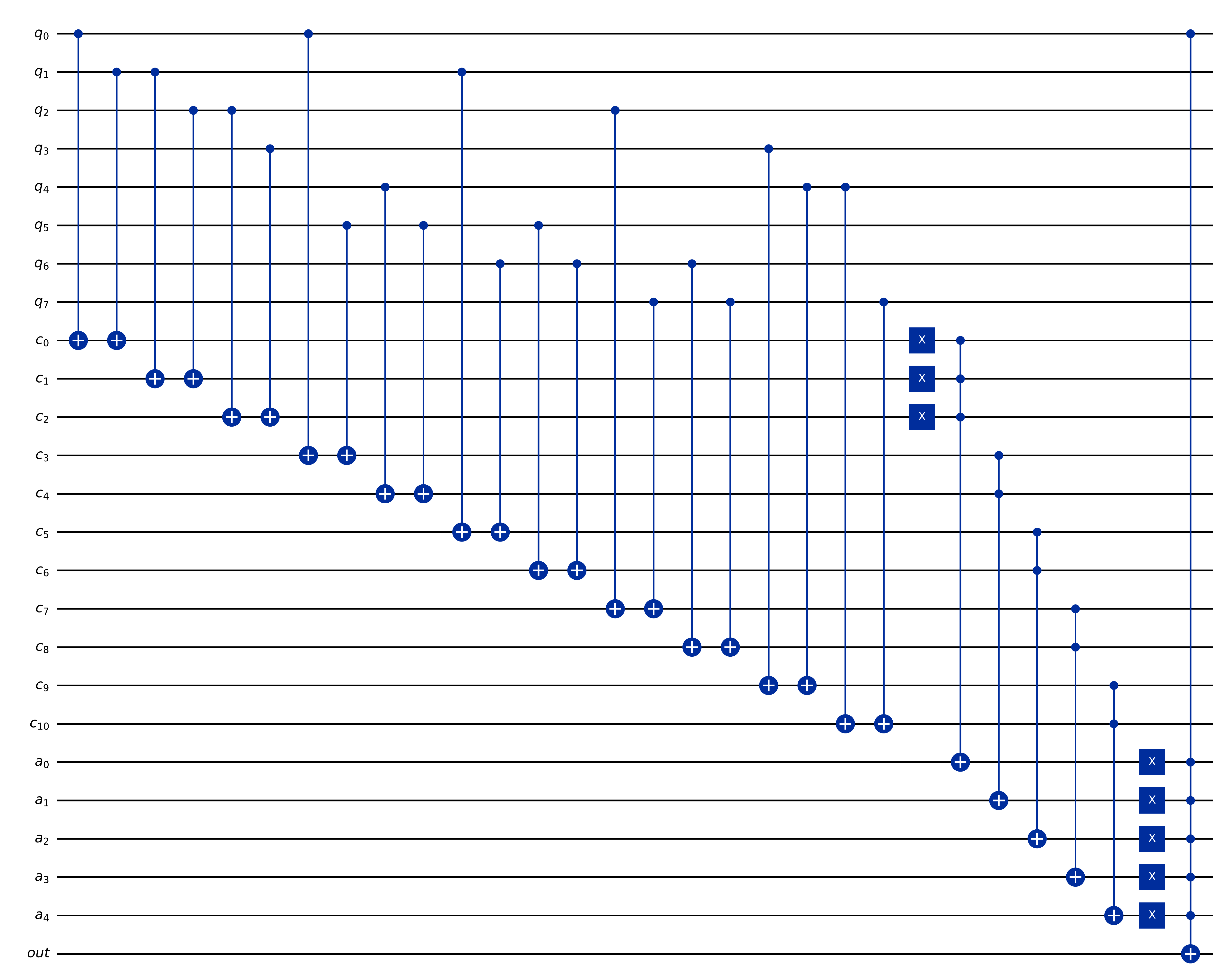}
				\subcaption[a]{N$^3$MLT}
			\end{minipage}
			\begin{minipage}{0.49\textwidth}
				\centering
				\includegraphics[width=\textwidth]{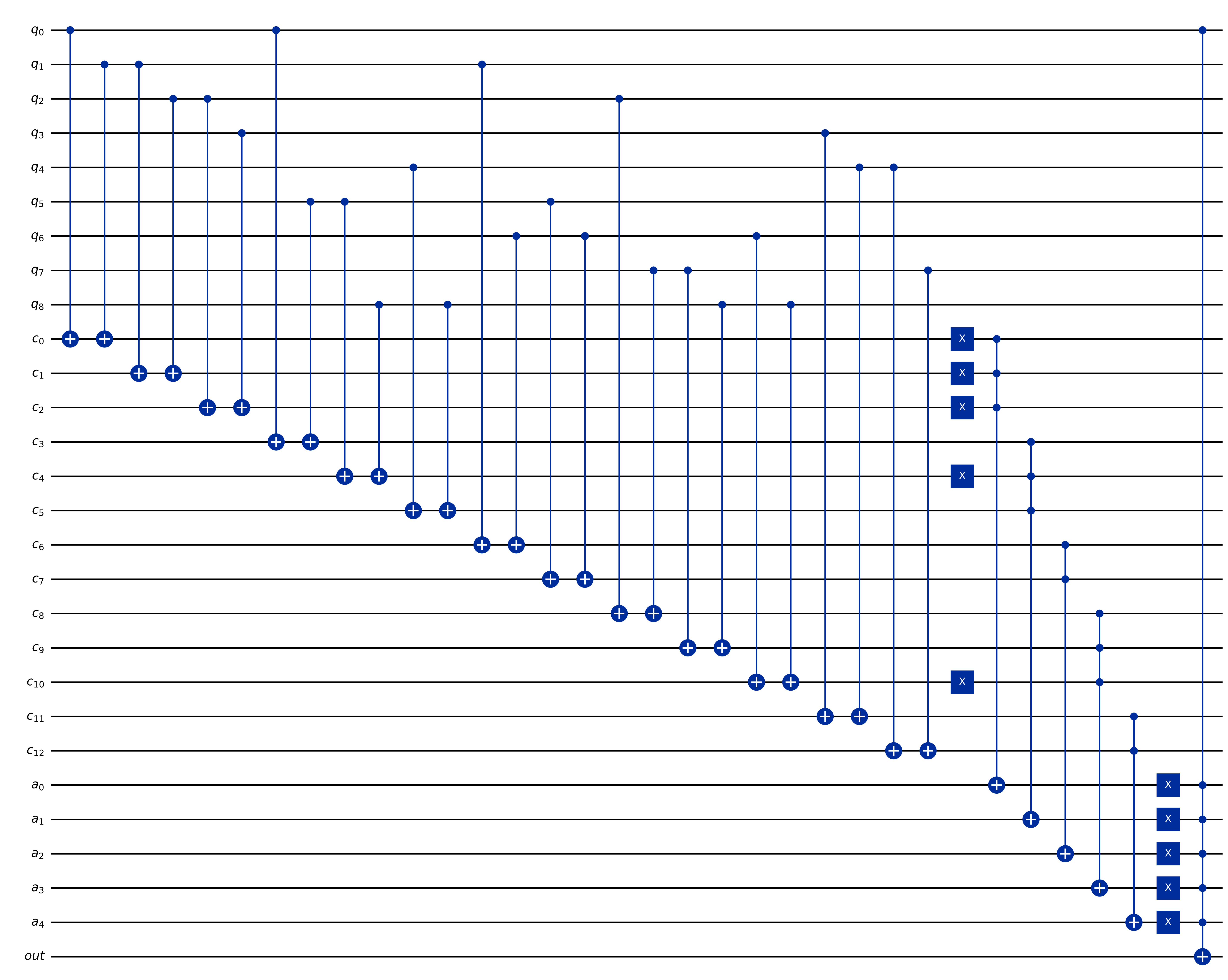}
				\subcaption[b]{$t$-channel}
			\end{minipage}\\
			\begin{minipage}{0.49\textwidth}
				\centering
				\includegraphics[width=\textwidth]{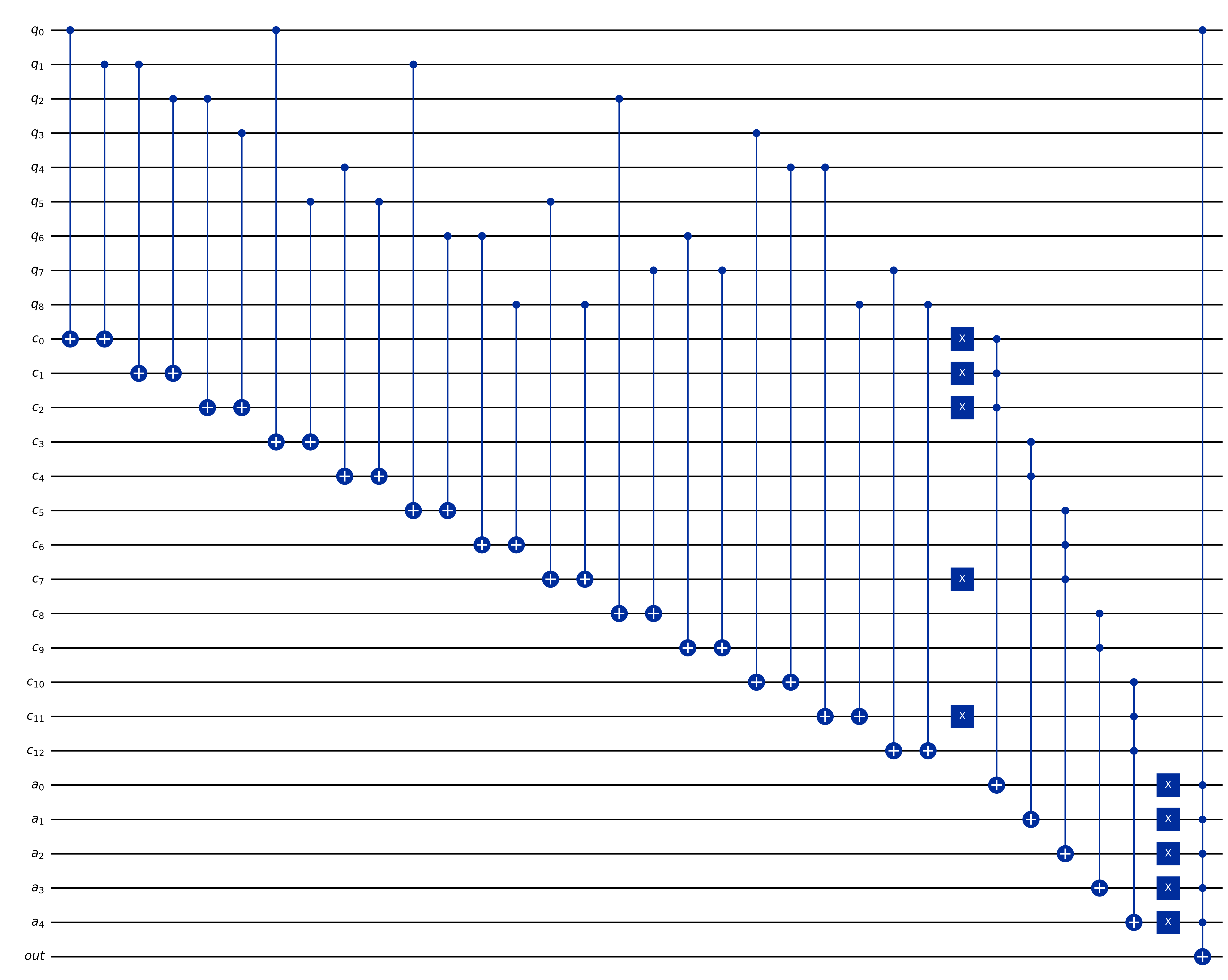}
				\subcaption[c]{$s$-channel}
			\end{minipage}
			\begin{minipage}{0.49\textwidth}
				\centering
				\includegraphics[width=\textwidth]{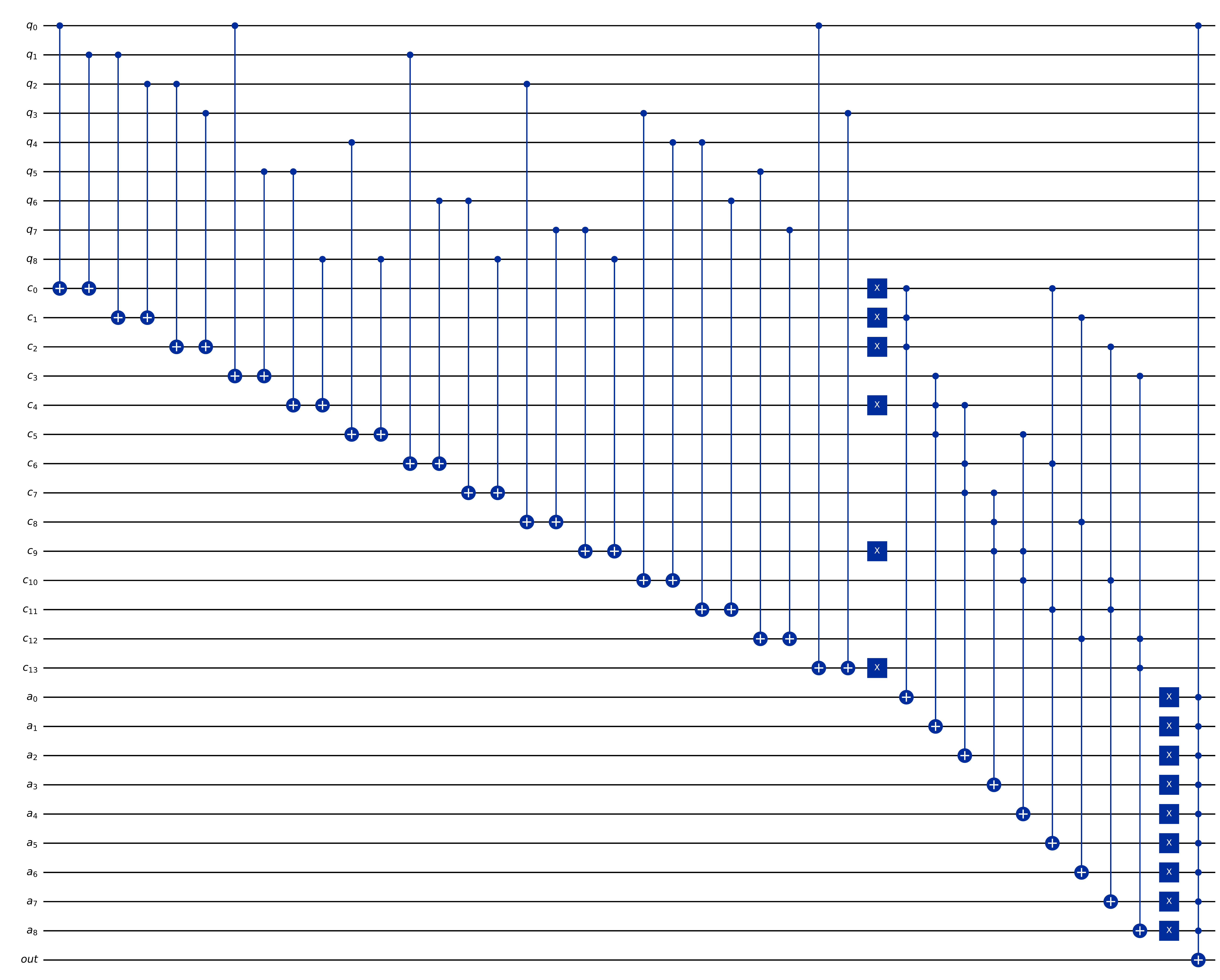}
				\subcaption[d]{$u$-channel}
			\end{minipage}

			\caption{Oracles of the quantum circuits for four-eloop topologies (omitting the reflection of the quantum gates).
            (a) N$^3$MLT, (b) $t$-channel, (c) $s$-channels and (d) $u$-channel of N$^4$MLT. 
            \label{fig:qc4eloops}}
		\end{figure}

\subsection{Four eloops}
\label{ssec:4eloops}
\begin{figure}[ht]
\begin{center}
\includegraphics[width=.96\textwidth, trim={5cm 0 4.5cm 1cm},clip]{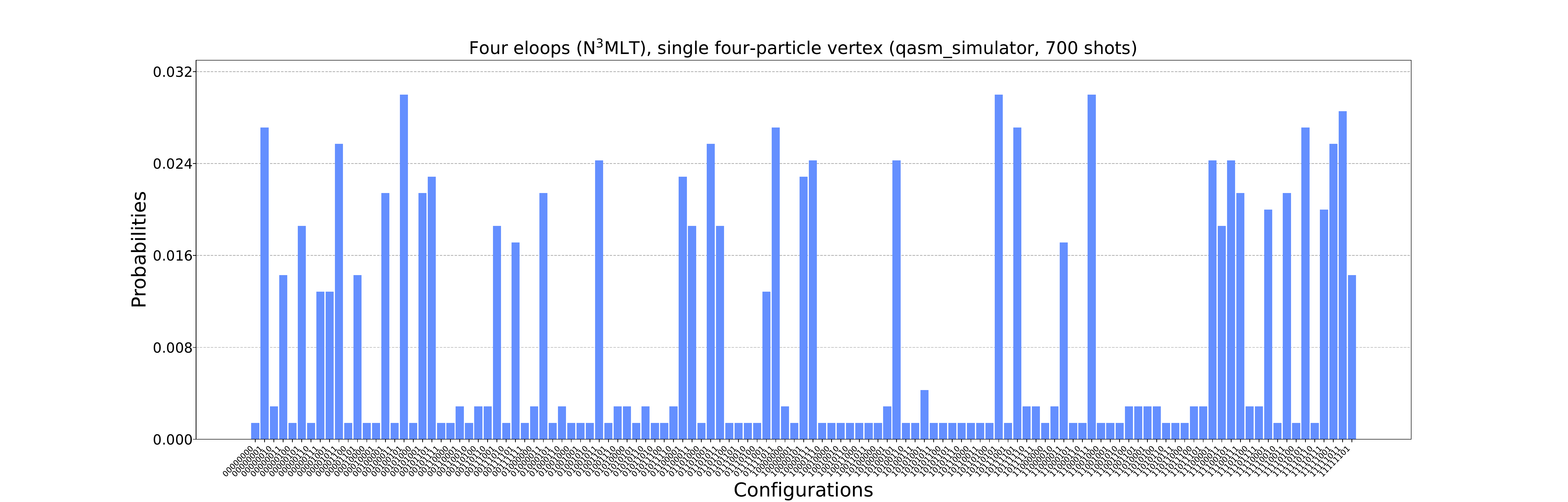} \\
\includegraphics[width=.96\textwidth, trim={5cm 0 4.5cm 1cm},clip]{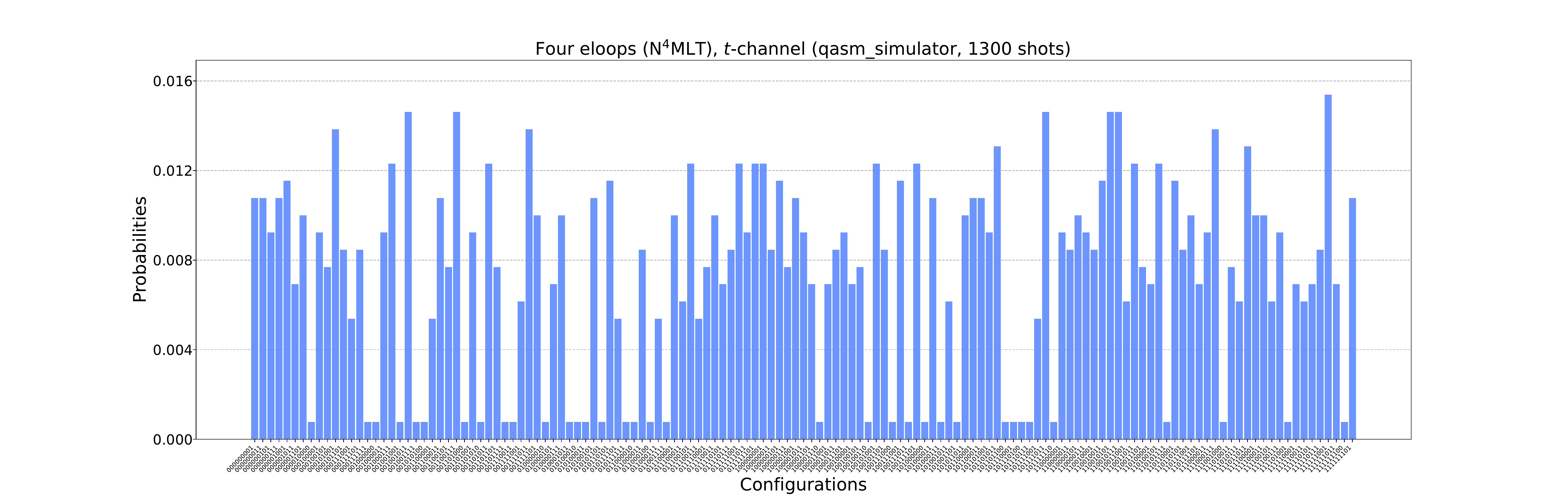}\\
\includegraphics[width=.96\textwidth, trim={5cm 0 4.5cm 1cm},clip]{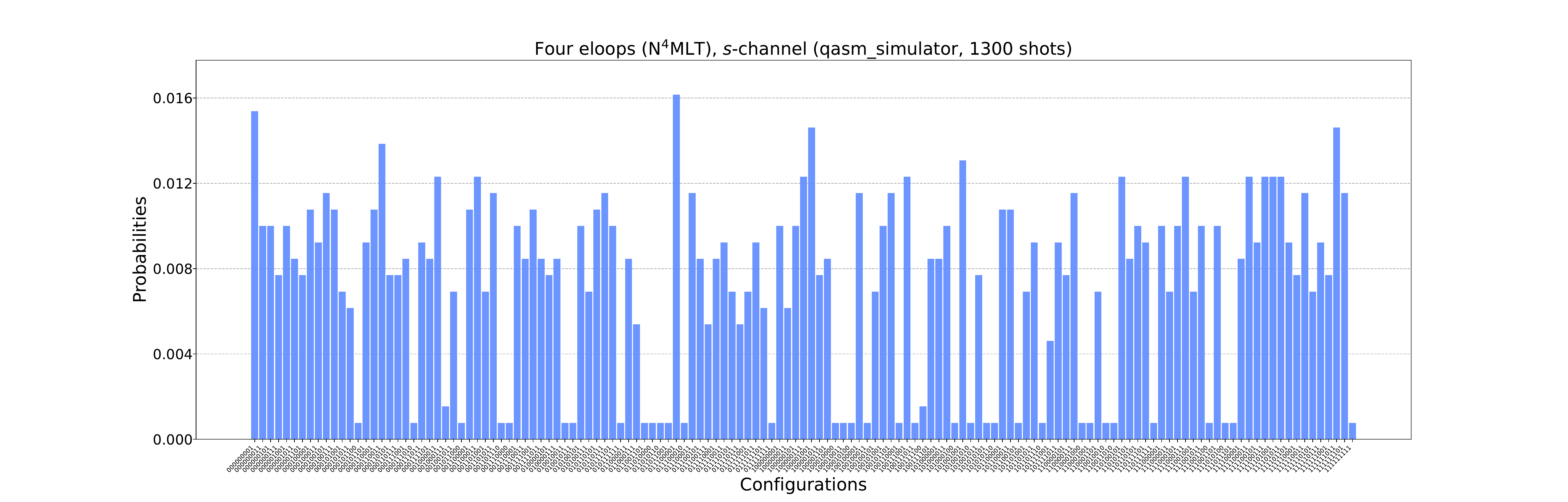} 
\end{center}
\caption{Probability distribution of causal and noncausal configurations for four-eloop topologies after 700, 1300
and 1300 shots with qasm\_simulator, respectively in the IBM's Qiskit framework. 
From top to down: N$^3$MLT, $t$-, and $s$-channels of N$^4$MLT with $n_i=1$. 
The number of selected states is $39/256$, $102/512$ and $102/512$, respectively.
\label{fig:prob4eloops}}
\end{figure}

\begin{figure}[ht]
    \centering
    \includegraphics[width=\textwidth, trim={5.25cm 0 6.2cm 1cm},clip]{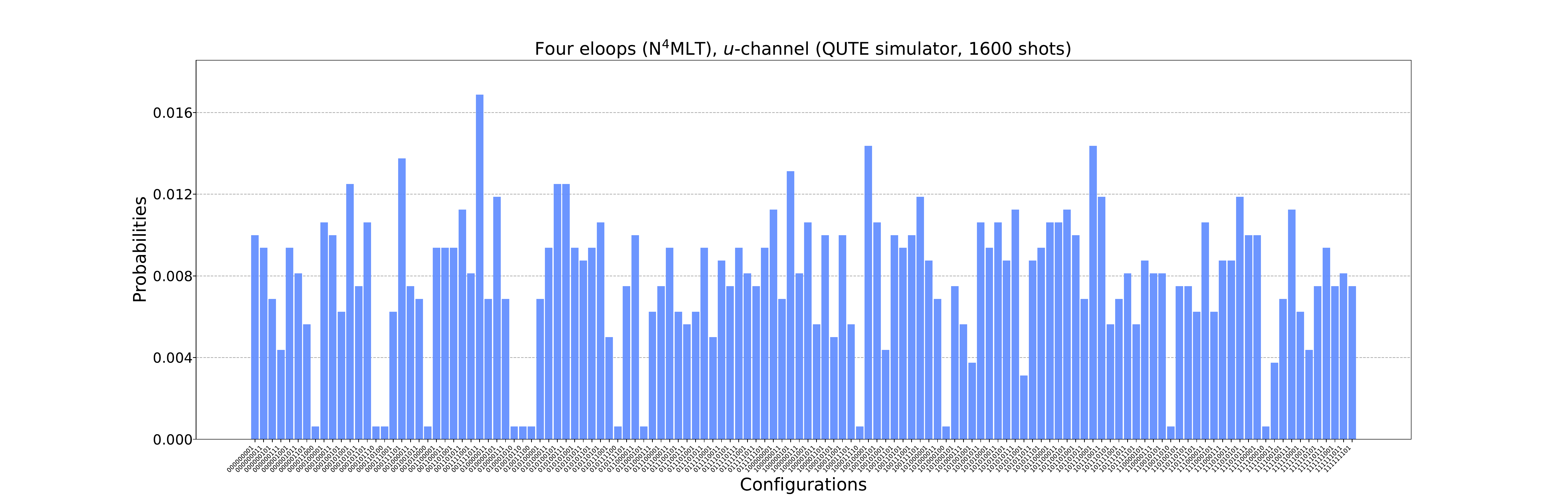}
    \caption{Probability distribution of causal and noncausal configurations for the four-eloop $u$-channel after 1600 shots in the QUTE Testbed framework. The number of selected states is $115/512$.}
    \label{fig:my_label}
\end{figure}

Starting at four loops, we should consider four different topologies (see Fig.~\ref{fig:multilooptopologies}D to \ref{fig:multilooptopologies}G).
The N$^3$MLT multiloop topology is characterized by 8 sets of edges connected through $5$ vertices. 
For $n_i=1$, with $i\in \{0, \ldots, 7\}$, the loop clauses are
\bea
&& a_0^{(4)} = \neg \left( c_{01} \wedge c_{12}\wedge c_{23} \right)~, \nn \\
&& a_1^{(4)} = \neg \left( \bar c_{05} \wedge \bar c_{45} \right)~,\nn \\
&& a_2^{(4)} = \neg \left( \bar c_{16} \wedge \bar c_{56} \right)~, \nn \\
&& a_3^{(4)} = \neg \left( \bar c_{27} \wedge \bar c_{67} \right)~, \nn \\
&& a_4^{(4)} = \neg \left( \bar c_{34} \wedge \bar c_{47} \right)~, \label{eq:4loopclauses}
\eea
and the Boolean test function is
\beq
f^{(4)}(a,q) = (a_0^{(4)} \wedge \ldots \wedge a_4^{(4)}) \wedge q_0~.
\eeq
Some of the loop clauses in \Eq{eq:4loopclauses} are common to the $t$-, $s$- and $u$-channels,
which are inclusively denoted as N$^4$MLT as they involve each one extra set of edges with respect
to N$^3$MLT. The channel specific loop clauses that are needed are 
\bea
&& a_1^{(t)} = \neg \left( \bar c_{05} \wedge c_{58} \wedge \bar c_{48} \right)~, 
\qquad a_3^{(t)} = \neg \left( \bar c_{27} \wedge \bar c_{78}  \wedge c_{68} \right)~, 
\eea 
\bea
&& a_2^{(s)} = \neg \left( \bar c_{16} \wedge \bar c_{68} \wedge c_{58} \right)~, 
\qquad a_4^{(s)} = \neg \left( \bar c_{34} \wedge c_{48}  \wedge \bar c_{78} \right)~,
\eea
and 
\bea
&& a_3^{(u)} = \neg \left( \bar c_{27} \wedge c_{78} \wedge \bar c_{68} \right)~,  \nn \\
&& a_4^{(u)} = \neg \left( \bar c_{34} \wedge \bar c_{48}  \wedge c_{78} \right)~, \nn \\
&& a_5^{(u)} = \neg \left( c_{01} \wedge \bar c_{16}  \wedge \bar c_{46} \right)~, \nn \\
&& a_6^{(u)} = \neg \left( c_{12} \wedge \bar c_{27}  \wedge \bar c_{57} \right)~, \nn \\
&& a_7^{(u)} = \neg \left( c_{23} \wedge \bar c_{34}  \wedge \bar c_{46} \right)~, \nn \\
&& a_8^{(u)} = \neg \left( c_{03} \wedge \bar c_{05}  \wedge \bar c_{57} \right)~.
\eea
The number of loop clauses for the $u$-channel is much larger than for the other configurations
because it is the first nonplanar diagram. 
Each of the $t$-, $s$- and $u$-channel is characterized by one of the following Boolean conditions
\bea
&& f^{(4,t)}(a,q) = \left(a_0^{(4)} \wedge a_1^{(t)} \wedge a_2^{(4)} \wedge a_3^{(t)} \wedge a_4^{(4)} \right) \wedge q_0~, \nn \\
&& f^{(4,s)}(a,q) = \left(a_0^{(4)} \wedge a_1^{(4)} \wedge a_2^{(s)} \wedge a_3^{(4)} \wedge a_4^{(s)} \right) \wedge q_0~, \nn \\
&& f^{(4,u)}(a,q) = \left(a_0^{(4)} \wedge a_1^{(t)} \wedge a_2^{(s)} \wedge a_3^{(u)} \wedge \ldots \wedge a_8^{(u)} \right) \wedge q_0~.
\eea
The number of qubits required for each configuration is given by the lower ranges of Tab.~\ref{tb:qubits}, 
i.e. $25$, $28$, $28$ and $33$ qubits respectively. Despite the complexity of these topologies, 
the quantum algorithm is well supported by the capacity of the IBM Quantum simulator (see Fig.~\ref{fig:prob4eloops}),
with the exception of the $u$-channel that was tested within the QUTE Testbed framework
as it supports more than 32 qubits (see Fig.~\ref{fig:my_label}).
Following the procedure described for three-eloop topologies, more complex topologies with $n_i \ge 1$ 
are also amenable to the quantum algorithm, although they may soon exceed the current capacity of the quantum simulator.
Representative bootstrapped diagrams at four eloops are shown in Fig.~\ref{fig:4eloops_th}.
The corresponding oracles of the quantum circuits are presented in Fig.~\ref{fig:qc4eloops}.

\subsection{Counting of causal states}
\label{ssec:Counting}
After discussing the causal structure of multiloop Feynman diagrams, it is clear that detecting all the 
configurations with causal-compatible momenta flow is crucial to identify the terms involved in the LTD 
representation of Eq. (\ref{eq:AD}). Also, as explained in Sec. \ref{sec:QuantumGrover}, the performance 
of quantum search algorithms depends on the number of winning states compared to all the possible configurations 
of the system. Thus, in this section, we present a counting of states fulfilling causality conditions for 
different topologies.

Given a reduced Feynman graph made of $V$ vertices connected through $n$ edges, there are $N=2^n$ possible 
orientations of the internal edges but only some of them are compatible with causality. Since causal-compatible 
momentum flows are in a one-to-one correspondence with the number of directed acyclic graphs, $n_A$, built 
from the original reduced Feynman graph, we are interested in estimating the ratio $r_A=n_A/2^n$. In order to 
do so, let us consider two extreme topologies:
\begin{itemize}
 \item \emph{Maximally Connected Graph} (MCG) where all the vertices are connected to each other and $n=V(V-1)/2$. 
 \item \emph{Minimally Connected Graph} (mCG) with $n=V$, i.e. the minimal number of edges. It only occurs at 
 one eloop.
\end{itemize}
For a fixed number of vertices $V$, the number of causal-compatible orientations is minimal for MCG and maximal 
for mCG; the bigger the number of edges, the larger the set of constraints that a graph must fulfil to be free 
of cycles. In fact, it is easy to show that
\beq
{V!} \leq n_A \leq 2^{(V-1)V/2} - 2 \, , 
\label{eqCOUNTINGLIMITS}
\eeq
which implies that, in the limit $V \to \infty$, we have $r_A \to 0$ for highly-connected topologies. In other 
words, for diagrams with a high number of eloops, the ratio $r_A$ is generally small. On the other hand, 
for one eloop diagrams, $r_A \to 1$ as $V \to \infty$, which means that the ratio of \emph{noncausal} flow 
configurations is small compared to the total number of configurations. In Tab.~\ref{tb:states} we present 
explicit values for the topologies described in Fig.~\ref{fig:multilooptopologies}, focusing on the number of 
causal configurations ($n_A$).

\begin{table}
\begin{center}
\begin{tabular}{cccc} \hline \hline 
Diagram & Vertices/Edges & $n_\lambda$  & $n_A/2^n$\\ \hline
one eloop ($n=3$) & 3/3 & 3 & 6/8\\
two eloops ($n_0=n_2=2,  n_1=1$) & 4/5 & 6& 18/32\\
three eloops ($n_i=1$)  & 4/6 & 7 & 24/64\\
N$^3$MLT ($n_i=1$)& 5/8 & 13 & 78/256\\
N$^4$MLT $t$-channel ($n_i=1$) & 6/9 & 22 & 204/512\\
N$^4$MLT $s$-channel ($n_i=1$)& 6/9 & 22 & 204/512\\
N$^4$MLT $u$-channel ($n_i=1$)& 6/9 & 24 & 230/512\\
\hline \hline 
\end{tabular}
\end{center}
\caption{Number of causal propagators ($n_\lambda$) and causal or acyclic configurations ($n_A$) 
for the topologies drawn in Figs. \ref{fig:multilooptopologies} and \ref{fig:causalmlt5}.}
\label{tb:states}
\end{table}

With these results on sight, we notice that $n_A$ turns out to become a small fraction of the total number of
flux-orientation for increasingly complex diagrams. Thus, we expect that the quantum search algorithms perform 
better against classical algorithms, since the number of 
winning states is a tiny fraction of the total space of states. On the other hand, when the number of noncausal 
configurations is small compared with the total number of states, we can revert the definition of Grover's marker 
and look for \emph{cyclic graphs} (i.e. noncausal flow configurations). For instance, for one-eloop topologies, 
we find that the ratio of noncausal versus total configurations is given by $2^{1-V(V-1)/2}$ and the 
original version of Grover's algorithm perfectly applies to this problem.

\section{Conclusions}
\label{sec:Conclusion}
We have presented the first proof-of-concept application of a quantum algorithm to multiloop Feynman integrals exploiting the loop-tree duality and causality. The specific problem we have addressed is the identification of all the causal singular configurations of the loop integrand resulting from setting on shell internal Feynman propagators. This information is useful both for identifying the physical discontinuities of the Feynman loop integral and to bootstrap its causal representation in the loop-tree duality. Beyond particle physics, this is also a challenging problem of identifying directed acyclic graphs.

We have described the quantum algorithm in general terms, and have provided the particular details on its implementation to selected multiloop topologies.
These cases were successfully handled by IBM Quantum and QUTE simulators. Even if for these selected topologies the quantum speed-up is attenuated by the number of shots required to identify all the causal configurations, more involved topologies and the selection of configurations satisfying further causality conditions would fully benefit from Grover's quadratic speed-up.

Given the quantum depth of the algorithm, its execution in
current real devices leads to unreliable results due to the present hardware limitations.
However, the quantum simulators successfully identifies 
all causal states even for the most complex multiloop configurations considered.


\section*{Acknowledgements}
We are very grateful to A. P\'erez for suggesting us to contact Fundaci\'on Centro Tecnol\'ogico de la Informaci\'on y la Comunicaci\'on (CTIC), and CTIC for granting us access to their simulator \emph{Quantum Testbed} (QUTE). We thank also access to IBMQ.
This work is supported by the Spanish Government (Agencia Estatal de Investigaci\'on)  and ERDF
funds from European Commission (Grant No. PID2020-114473GB-I00),  Generalitat  Valenciana
(Grant No. PROMETEO/2021/071) and the  COST  Action  CA16201  PARTICLEFACE.  
SRU acknowledges support from CONACyT and  Universidad  Aut\'onoma  de  Sinaloa;  
AERO from the Spanish Government (PRE2018-085925).
LVS acknowledges funding from the European Union’s Horizon 2020 research and innovation
programme under the Marie Sklodowska-Curie grant agreement No 101031558.

\bibliographystyle{JHEP}

\providecommand{\href}[2]{#2}\begingroup\raggedright\endgroup

\ULforem
\end{document}